\documentclass[aps,prc,superscriptaddress,twocolumn,10pt]{revtex4-1}
\usepackage{aas_macros}
\usepackage{graphicx}
\usepackage{txfonts}
\usepackage{dcolumn}
\newcolumntype{d}[1]{D{.}{.}{#1}}
\newcommand{\tabhead}[1]{\multicolumn{1}{c}{#1}}

\newcommand{\src}{KS~1731-260}
\newcommand{\mxb}{MXB~1659-29}
\newcommand{\xtej}{XTE~J1701-462}
\newcommand{\MeV}{{\rm ~MeV }}

\newcommand{\ergpersec}{\ensuremath{\mathrm{erg\,s^{-1}}}}

\newcommand{\Tc}{\tilde{T}}
\newcommand{\Teff}{\ensuremath{T_{\!\mathrm{eff}}}}
\newcommand{\Teffinf}{\ensuremath{\Teff^\infty}}
\newcommand{\Qnuc}{\ensuremath{Q_{\mathrm{nuc}}}}
\newcommand{\Qimp}{\ensuremath{Q_{\mathrm{imp}}}}
\newcommand{\Lph}{\ensuremath{L_\gamma}}
\newcommand{\Lnu}{\ensuremath{L_\nu}}
\newcommand{\Lin}{\ensuremath{L_{\mathrm{in}}}}
\newcommand{\epsdUrca}{\ensuremath{\epsilon_\nu^{\mathrm{dU}}}}
\newcommand{\epsmUrca}{\ensuremath{\epsilon_\nu^{\mathrm{mU}}}}
\newcommand{\LnudUrca}{\ensuremath{L_{\nu,\mathrm{dU}}}}
\newcommand{\LnumUrca}{\ensuremath{L_{\nu,\mathrm{mU}}}}
\newcommand{\trec}{\ensuremath{t_{\mathrm{r}}}}
\newcommand{\tout}{\ensuremath{t_{\mathrm{o}}}}
\newcommand{\amu}{\ensuremath{m_{\mathrm{u}}}}

\usepackage[usenames]{xcolor}

\begin{document}

\title{A lower limit on the heat capacity of the neutron star core}

\author{Andrew Cumming}
\email{andrew.cumming@mcgill.ca}
\affiliation{Department of Physics and McGill Space Institute, McGill University, 3600 rue University, Montreal QC, Canada H3A 2T8}

\author{Edward F. Brown}
\email{ebrown@pa.msu.edu}
\affiliation{Department of Physics and Astronomy, Michigan State University, 567 Wilson Rd, East Lansing, MI 48864, USA}

\author{Farrukh J. Fattoyev}
\email{ffattoye@indiana.edu}
\author{C.~J.~Horowitz}
\email{horowit@indiana.edu}
\affiliation{Center for Exploration of Energy and Matter and Department of Physics, Indiana University, Bloomington, IN 47405, USA}

\author{Dany Page}
\email{page@astro.unam.mx}
\affiliation{Instituto de Astronom\'ia, Universidad Nacional Aut\'onoma de M\'exico, M\'exico, D.F. 04510, M\'exico}

\author{Sanjay Reddy}
\email{sareddy@uw.edu}
\affiliation{Institute for Nuclear Theory, University of Washington, Seattle, WA 98195,USA}

\date{\today}

\begin{abstract}
We show that observations of the core temperature of transiently-accreting neutron stars combined with observations of an accretion outburst give a lower limit to the neutron star core heat capacity. For the neutron stars in the low mass X-ray binaries \src, \mxb, and \xtej, we show that the lower limit is a factor of a few below the core heat capacity expected if neutrons and protons in the core are paired, so that electrons provide the dominant contribution to the heat capacity. This limit rules out a core dominated by a quark color-flavor-locked (CFL) phase, which would have a much lower heat capacity. Future observations of or limits on cooling during quiescence will further constrain the core heat capacity.
\end{abstract}

\maketitle

\section{Introduction}\label{s.introduction}

The next few years promise new constraints on the dense matter inside neutron stars. Progress on the equation of state may come from the detection of gravitational waves from merging neutron stars \cite{Lackey2012,Chatziioannou2015}, discovery of a neutron star with mass above $2\ M_\odot$ \cite{Antoniadis2016}, a measurement of the neutron star moment of inertia using the double pulsar system \cite{Lattimer2005,Kramer2009}, or precise mass and radius determination by X-ray timing with the Neutron Star Interior Composition Explorer (NICER) \cite{Watts2016}. Also of great interest are observations of neutron star thermal evolution, in particular for neutron stars in close binaries that undergo transient accretion outbursts. Observations of the temperatures of these neutron stars have been used to limit the neutrino emissivity of the neutron star core \cite{Fujimoto1984,Colpi:2001,Yakovlev2004}, which sensitively depends on the composition. Most recently, it has been shown that long-term monitoring of neutron stars after accretion outbursts provides information on neutron star crust physics. How the neutron star cools over months to years after accretion ends depends on the thermal conductivity and heat capacity of the neutron star crust \cite{Shternin2007a,Brown2009,Page2013,Turlione2015}, including the pasta regions near nuclear density \cite{Horowitz2015}. In this paper, we show that continued observations of these systems on timescales of years can be used to obtain the first limits on the heat capacity of the neutron star core.

Accreting neutron stars are useful sources to study thermal evolution because their crusts are driven out of thermal equilibrium by heating from compression-driven nuclear reactions \cite{Haensel1990,Haensel2008,Brown1998}. 
The crust reactions release an energy $\Qnuc\approx 1$--$2\ {\rm MeV}/\amu$ per accreted nucleon \cite{Haensel1990,Haensel2008}, $\amu$ being the atomic mass unit; most of this heat is conducted inwards to the core. The luminosity entering the core is $\Lin\approx \dot M \Qnuc$, where $\dot M$ is the mass accretion rate.
The thermal conductivity in the core is large enough that it remains isothermal, described by a redshifted temperature $\Tc$ (so that the local temperature at radius $r$ is $T(r)=\Tc e^{-\phi(r)}$ where $\phi(r)$ is the gravitational potential \cite{Glen1980}). The thermal evolution of the neutron star core during outburst is given by \citep[e.g.,][]{Page2004}
\begin{equation}\label{eq:evolution}
C{d\Tc\over dt} = -\Lnu + \Lin,
\end{equation}
where $C$ is the heat capacity, and $\Lnu$ is the core neutrino cooling luminosity. Both $\Lnu$ and $\Lin$ are defined in the frame of infinitely distant observer; the connection between $C$ and $L$ with local quantities is given by relativistic stellar structure equations \citep[see, e.g.,][]{thorne77}.

Over long timescales, the core temperature reaches a value at which the heating is balanced by neutrino losses and radiative losses during quiescence
\begin{equation}\label{eq:Lbalance}
\Lnu(\Tc) + \Lph(\Tc) \approx  \langle\dot M\rangle \Qnuc ,
\end{equation}
where $\langle \dot M\rangle$ is the accretion rate averaged over outburst and quiescent periods \cite{Brown1998}. Observations of more than twenty low mass X-ray binaries have measured the surface luminosity of the neutron star in quiescence, providing a measurement of the core temperature \cite{Heinke2009}. Comparing the measured luminosities to theoretical predictions determines the neutrino emissivity of the neutron star core. The observations show a range of quiescent temperatures, compatible with modified Urca emissivity in some sources, and indications of enhanced emissivity in others \cite{Beznogov2015a,Beznogov2015b}.
How long a time span is needed to obtain a valid average $\langle \dot M \rangle$ is given by the Kelvin-Helmholtz time-scale of the star, which depends on the heat capacity and ranges from $10^2$ to $10^5$ years \citep{Wijnands_DP:2013}.

This previous work assumes that the core has reached its equilibrium temperature, and that the energy released during any single accretion outburst does not significantly change the core temperature. In this paper, we consider the opposite limit. For cold neutron stars, the core heat capacity is rather low, allowing the core to heat up significantly during outburst and cool down in quiescence. During an outburst of duration $\tout$, an energy $E\approx \dot M \Qnuc \tout$ flows into the core from the crust. Assuming that the core heat capacity $C$ is proportional to temperature (as appropriate for degenerate fermions), $C=A\Tc$ where $A$ is a constant, and that neutrino losses are negligible, the core will increase in temperature by an amount $\Delta \Tc=\Tc_f-\Tc_i$ given by ${\Tc_f}^2-{\Tc_i}^2=2E/A$, where $\Tc_f$ and $\Tc_i$ are the initial and final temperatures. If the core starts off cold, $\Tc_i \ll \Tc_f$, then $\Delta \Tc\approx \Tc_f$. Solving for the heat capacity, we then find $C=A\Tc_f = 2E/\Tc_f$, which can be calculated given values for $\dot M$, $\tout$, and $\Tc_f$ obtained from observations. In reality, we do not know the starting temperature of the core and so this is actually a lower limit on the heat capacity 
\begin{equation}\label{eq:Clim_intro}
C>\frac{2E}{\Tc_f},
\end{equation}
where larger values of $C$ mean that the core temperature started off closer to the measured value after the outburst.

To illustrate that this might be an interesting limit, consider an outburst with accretion rate of $\dot M\approx 0.1\ \dot M_{\rm Edd}\approx 10^{17}\ {\rm g\ s^{-1}}$ (where $\dot M_{\rm Edd}$ is the Eddington accretion rate) and duration 10 years. The energy deposited into the core is then
\begin{equation}\label{eq:E}
E\approx 6.0\times 10^{43}\ {\rm erg}\ \left(\frac{\dot M}{10^{17}\ {\rm g\ s^{-1}}}\right)\left(\frac{\tout}{10\ {\rm yr}}\right)\left(\frac{\Qnuc}{2\ {\rm MeV}/\amu}\right).
\end{equation}
A core temperature of $\Tc=10^8\ {\rm K}$ then gives a limit $C\gtrsim 10^{36}\,\Tc_8\ {\rm erg\ K^{-1}}$ from Eq.~(\ref{eq:Clim_intro}), where we use the shorthand notation $\Tc_8\equiv \Tc/10^8\,\mathrm{K}$. For a core consisting of non-superfluid neutrons and protons, the heat capacity is expected to be $C\sim 10^{38}\,\Tc_8\ {\rm erg\ K^{-1}}$ (e.g.~\cite{Page1994}), much larger than the limit. However, the nucleons may be superfluid, in which case their contribution to the specific heat is suppressed exponentially, and the heat capacity is set by the leptons giving a value $C\sim 10^{37}\,\Tc_8\ {\rm erg\ K^{-1}}$. Even smaller values of heat capacity are possible, for example if the high density matter forms a color-flavor-locked (CFL) phase \cite{Alford1999}, lowering the electron content of the core. 

We start in Sec.~\ref{s.limit-core-C} by looking in detail at the low mass X-ray binary \src, which was observed in outburst for over 12 years before going into quiescence in 2001 \cite{Wijnands2002}. The long outburst makes \src\ a promising source to derive the heat capacity limit, since it should have deposited the most energy into the core; it also has one of the lowest measured temperatures, and has been monitored for almost 15 years in quiescence \cite{Cackett2010,Merritt2016}. We use the quiescent temperature measurement of \src\ to constrain the neutron star core temperature, and model the thermal relaxation of the neutron star in quiescence to constrain the crust heating and envelope composition. We then discuss the theoretical expectations for the neutron star heat capacity and compare to the lower limit from \src\ and other sources (Sec.~\ref{s.comparison-with-predictions}). We discuss what further limits could be obtained on the core heat capacity by continued monitoring of these sources in future years (Sec.~\ref{s.core-evolution}).

\section{The limit on core heat capacity from \src}
\label{s.limit-core-C}

In this section, we investigate in more detail the lower limit on the core heat capacity from observations of \src. We discuss the core temperature (Sec.~II.A), use the cooling curve to deduce the envelope composition and investigate uncertainties in input parameters (Sec.~\ref{s.energy-deposited-into-core}), and the effect of accretion rate variations in outburst (Sec.~\ref{s.accretion-rate-variations}).

\subsection{Core temperature}
\label{s.core-temperature}

Following the outburst, the neutron star temperature in \src\ was observed to decline over 8 years from $\Teffinf=103\ {\rm eV}$ to $63.1\ {\rm eV}$ as the crust relaxed back into thermal equilibrium with the core \cite{Cackett2010}. 
Once the star has thermally relaxed most of its interior, i.e., its core and most of its crust, are isothermal except for the outermost layers, which are called the {\em envelope}. Models of the neutron star envelope \citep{Gudmundsson1983,Potekhin1997} and of the temperature gradient permeating it allow us to estimate a range of values for $\Tc$ from the observed $\Teffinf$.

A large uncertainty in determining $\Tc$ from the observations is the composition of the envelope. During the outburst, light elements accumulate on the surface of the star where they burn to heavy element ashes. Depending on the state of the burning at the end of the outburst, the envelope can consist of heavy or light elements (e.g.~\cite{Brown2002}). For an iron envelope, the relation between the effective temperature and core temperature for a thermally relaxed neutron star is $T_{c,8}=1.288\,(T_{s,6}^4/g_{14})^{0.455}$ \cite{Gudmundsson1983} where $T_{s,6} = (\Teffinf/10^6\ {\rm K}) (1+z)$, $T_c = \Tc(1+z)$, $1+z \equiv e^{-\phi(R)}$ being the surface redshift factor, and we write the surface gravity $g=(GM/R^2)(1+z)$ as $g_{14} = g/10^{14}\ {\rm cm\ s^{-2}}$. For a $1.4\ M_\odot$, $12\ {\rm km}$ neutron star, $g_{14}=1.6$ and $1+z=1.24$, and we find 
\begin{equation}\label{eq:Tc_heavy}
\Tc=7.0\times 10^7\ {\rm K}\ \left({\Teffinf\over 63.1\ {\rm eV}}\right)^{1.82}\hspace{0.5cm}\textrm{(Fe envelope)}.
\end{equation}
For a light element envelope, the core temperature will be lower, because the envelope is less opaque. In that case, $T_{c,8}=0.552\,(T_{s,6}^4/g_{14})^{0.413}$ \cite{Potekhin1997}, and we find \begin{equation}\label{eq:Tc_light}
\Tc=3.1\times 10^7\ {\rm K}\ \left({\Teffinf\over 63.1\ {\rm eV}}\right)^{1.65}\hspace{0.5cm}\textrm{(He envelope)}.
\end{equation}
A different choice of mass and radius does not change the inferred core temperature dramatically. For example, for a much more compact neutron star with mass $2M_\odot$ and radius $10\ {\rm km}$, giving $g_{14}=4.18$ and $1+z=1.57$, we find core temperatures about 20\% smaller: $2.4\times 10^7\ {\rm K}$ for a light element envelope and $5.5\times 10^7\ {\rm K}$ for a heavy element envelope. Our value of core temperature is in good agreement with \cite{Ootes2016} who also used a heavy element envelope (column depth of helium $y_{He}=10^6\ {\rm g\ cm^{-2}}$) and found $\Tc=6.6\times 10^7\ {\rm K}$ (Eq.~[\ref{eq:Tc_heavy}] gives $6.4\times 10^7\ {\rm K}$ for their choice of $M=1.5\ M_\odot$ and $R=11\ {\rm km}$).

\subsection{Energy deposited in the core and the limit on heat capacity}
\label{s.energy-deposited-into-core}

To improve on Eq.~(\ref{eq:E}) and further constrain the energy that was deposited in the neutron star core, we carried out crust cooling simulations following \cite{Brown2009} to model the observed decline of the temperature of \src. The adjustable parameters in the model are the core temperature, the impurity parameter $\Qimp$, which sets the thermal conductivity of the inner crust (see \cite{Roggero2016} for a recent discussion of the interpretation of $\Qimp$), the temperature at the top of the grid $T_b$ at a density $\rho\approx 6\times 10^8\ {\rm g\ cm^{-3}}$, the mass and radius of the neutron star, the accretion rate during the outburst, and the composition of the neutron star envelope. As discussed by \cite{Brown2009}, the role of $\dot M$ is to determine the strength of the deep crustal heating $\dot M \Qnuc$, where we assume $\Qnuc=1.7\ {\rm MeV}/\amu$ \cite{Haensel1990,Haensel2008,Gupta2007}, while $T_b$ determines the temperature in the outer crust set by shallow heat sources (see \cite{Ootes2016} for a recent discussion of shallow heat sources). 

We consider two different envelope compositions which we refer to as Fe (iron) and He (helium) envelopes. The He envelope is the same as in \cite{Brown2009} and has a column depth of helium of $10^9\ {\rm g\ cm^{-2}}$; the Fe envelope has a helium column depth $10^4\ {\rm g\ cm^{-2}}$ and a composition of iron at higher column depth. Although a continuum of envelope models is possible with different values of $y_{He}$, the effect of the envelope depends on the composition in a narrow range of densities known as the sensitivity strip \cite{Gudmundsson1983}, so in practice it is a good approximation to consider only the two cases in which either a light element or heavy element is present in the sensitivity strip region.

\begin{figure}[htbp]
\includegraphics[width=\columnwidth]{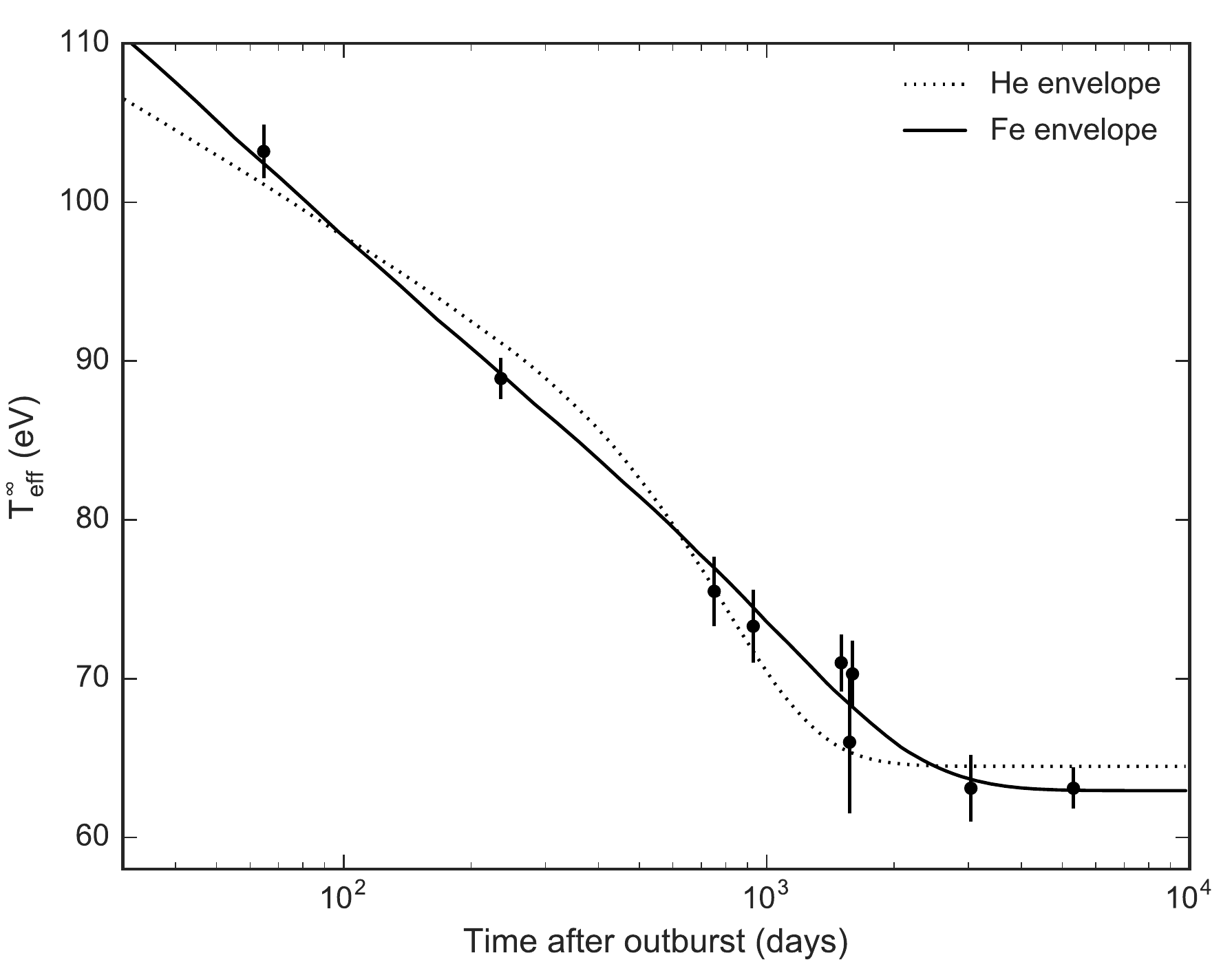}  
\caption{Example fits to the \src\ temperature measurements. The model shown as a solid line has an outburst duration of 12.0 years, accretion rate 0.1 Eddington, a heavy element envelope, $\Qimp=1.3$, a fixed $T_b=4.0\times 10^8\ {\rm K}$ at the top of the crust during the accretion phase, and a core temperature $\Tc=7.4\times 10^7\ {\rm K}$. The neutron star mass and radius are $1.4\ M_\odot$ and $12\ {\rm km}$. The inwards luminosity during the outburst reaches $2.4\times 10^{35}\ {\rm erg\ s^{-1}}$. The dotted curve is a model with a helium envelope as in \cite{Brown2009}, and has a core temperature of $\Tc=3.6\times 10^7\ {\rm K}$, while $\Qimp = 2.3$ and $T_b=2.2\times 10^8\ {\rm K}$. The inwards luminosity during the outburst reaches $2.2\times 10^{35}\ {\rm erg\ s^{-1}}$.}
\label{fig:tc}
\end{figure}

Two example lightcurves are shown in Figure \ref{fig:tc} compared to the observed temperatures for \src. The data points are from \cite{Cackett2010} with the addition of the latest temperature measurement \cite{Merritt2016}.  This latest measurement is consistent with the previous temperature measured 6 years earlier and
confirms that the star has wholly relaxed from the accretion outburst. 
For these models, we set $M=1.4\ M_\odot$, $R=12\ {\rm km}$, and $\dot M=1.3\times 10^{17}\ {\rm g\ s^{-1}}$, and then adjust $T_b$, $T_c$ and $\Qimp$ to obtain the best fit to the data. The model shown as a dotted curve uses a light element envelope and has similar parameters to \cite{Brown2009}. However, we find that a heavy element envelope gives a much better fit to the data (solid curve). This is in agreement with \cite{Ootes2016} who found a best fit column depth of light elements of $10^6\ {\rm g\ cm^{-2}}$. The main effect of the opaque heavy element envelope is to increase the crust temperature needed to match the data. The hotter outer crust has a lower thermal conductivity because of increased electron-phonon scattering. The steeper temperature gradient translates into faster cooling initially ($t\lesssim 200\ {\rm d}$) (see Eq.~[12] of Ref.~\cite{Brown2009}). However, in the inner crust, there is a reduced temperature contrast with the core and a larger heat capacity leading to slower cooling in the latter part of the lightcurve ($t\gtrsim 200\ {\rm d}$). The overall effect is to flatten out the break in the cooling curve compared to a light element envelope, bringing it into agreement with the data.

\begin{figure}[htbp]
\includegraphics[width=\columnwidth]{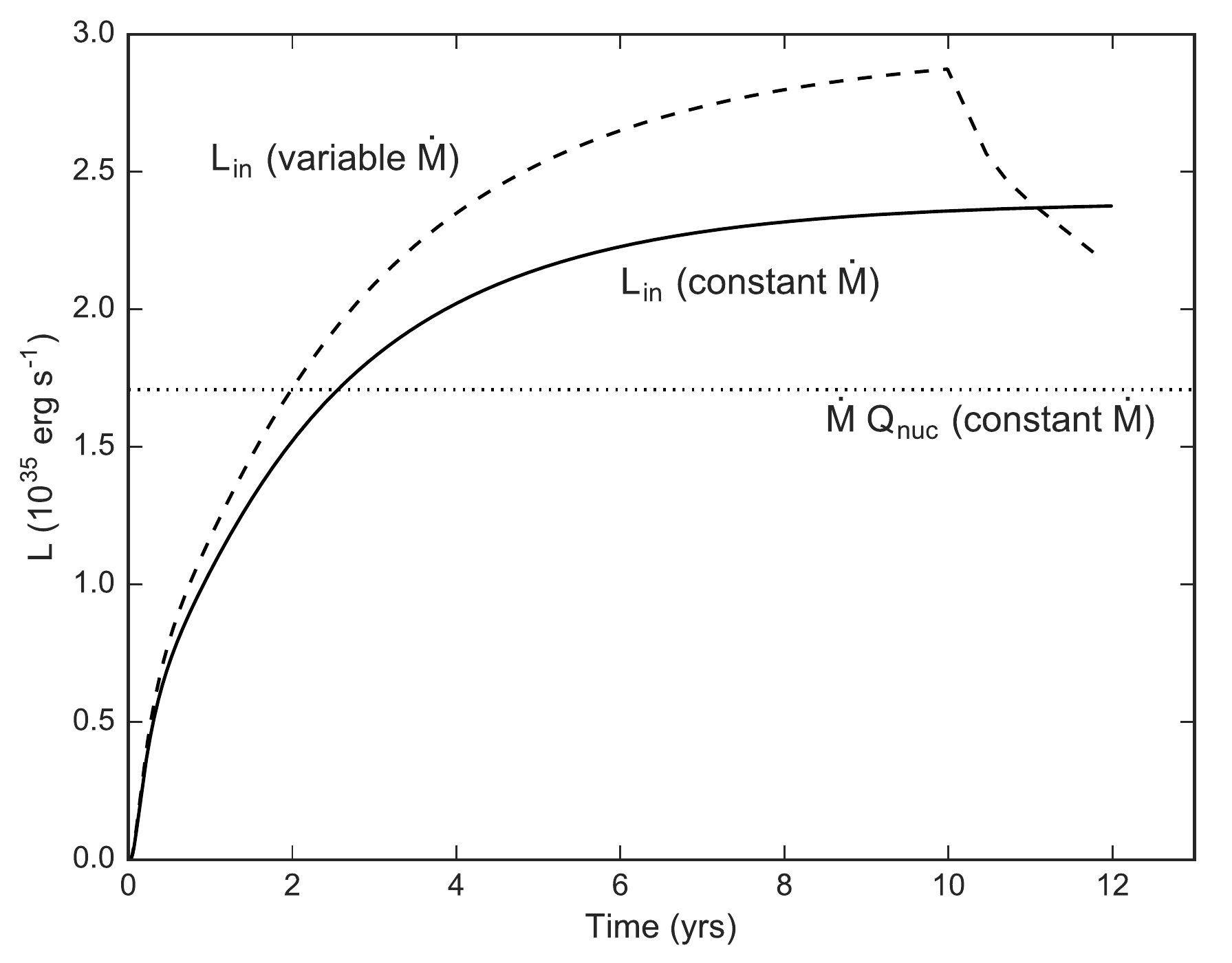}  
\caption{The luminosity entering the core $\Lin$ during the outburst, assuming either constant accretion rate or variable accretion rate. These models have a heavy element envelope. The dotted line shows the luminosity from crust reactions for the constant accretion rate case. The luminosity at late times is larger than this value because of the additional heat flowing into the crust from the upper boundary (representing shallow heat sources). All luminosities shown are for an observer at infinity.
\label{fig:Lin}}
\end{figure}

Figure \ref{fig:Lin} shows the luminosity entering the core during the outburst. During quiescence, there is an outwards luminosity at the crust/core boundary as the core cools. Once accretion begins, the crust is heated, and the luminosity at the core boundary changes sign and starts to increase in magnitude. After approximately 2000 days the crust is nearly in a thermal steady-state and the luminosity entering the core asymptotically approaches $\dot M\Qnuc+L_{\rm top}$, the sum of the energy released by nuclear reactions in the crust and the luminosity entering the crust from lower densities. For the heavy element envelope model, the inwards luminosity in the steady state is $2.4\times 10^{35}\ {\rm erg\ s^{-1}}$. The shallow heating from the top of the crust accounts for 30\% of the total, $6.9\times 10^{34}\ {\rm erg\ s^{-1}}$. Integrating over the outburst, we find a total energy $E = 7.5\times 10^{43}\ {\rm erg}$. The light envelope model has $E$ about 20\% smaller. Note that $E$ is defined in the frame of a distant observer, as are $Q$ and $\dot{M}$.

Combining the observed core temperature with the energy release during outburst gives the lower limit (from Eq.~[\ref{eq:Clim_intro}])
\begin{eqnarray}\label{eq:Climit2}
C&>&2.1\times 10^{36}\ {\rm erg\ K^{-1}}\ \left(\frac{\Tc_7}{7}\right)^{-1}\left(\frac{E}{7.5\times 10^{43}\ {\rm erg}}\right),
\label{eq:C_lowerbound}
\end{eqnarray}
where we use the value of $\Tc$ from the heavy element envelope since that gives the best fit to the cooling curve.  Using the assumed $C\propto T$ scaling gives
\begin{eqnarray}\label{eq:Climit}
{C\over \Tc_{8}}&>&3.1\times 10^{36}\ {\rm erg\ K^{-1}}\ \left(\frac{\Tc_{7}}{7}\right)^{-2}\left(\frac{E}{7.5\times 10^{43}\ {\rm erg}}\right).
\end{eqnarray}

We used a MCMC method to explore the parameter space to check the sensitivity of $E$ to changes in parameters. We used the \texttt{emcee} package \cite{ForemanMackey2013} to drive our cooling code. Rather than neutron star mass $M$ and radius $R$ as parameters, we use $R$ and surface gravity $g$. This helps convergence because mass and radius variations enter most directly into the cooling curves through the surface gravity, which sets the crust thickness (crust thickness $\propto 1/g^2$) \cite{Brown2009}. In all cases we use a heavy element envelope. We adopt uniform priors for all variables. 

\begin{figure*}[htbp]
\includegraphics[width=2.05\columnwidth]{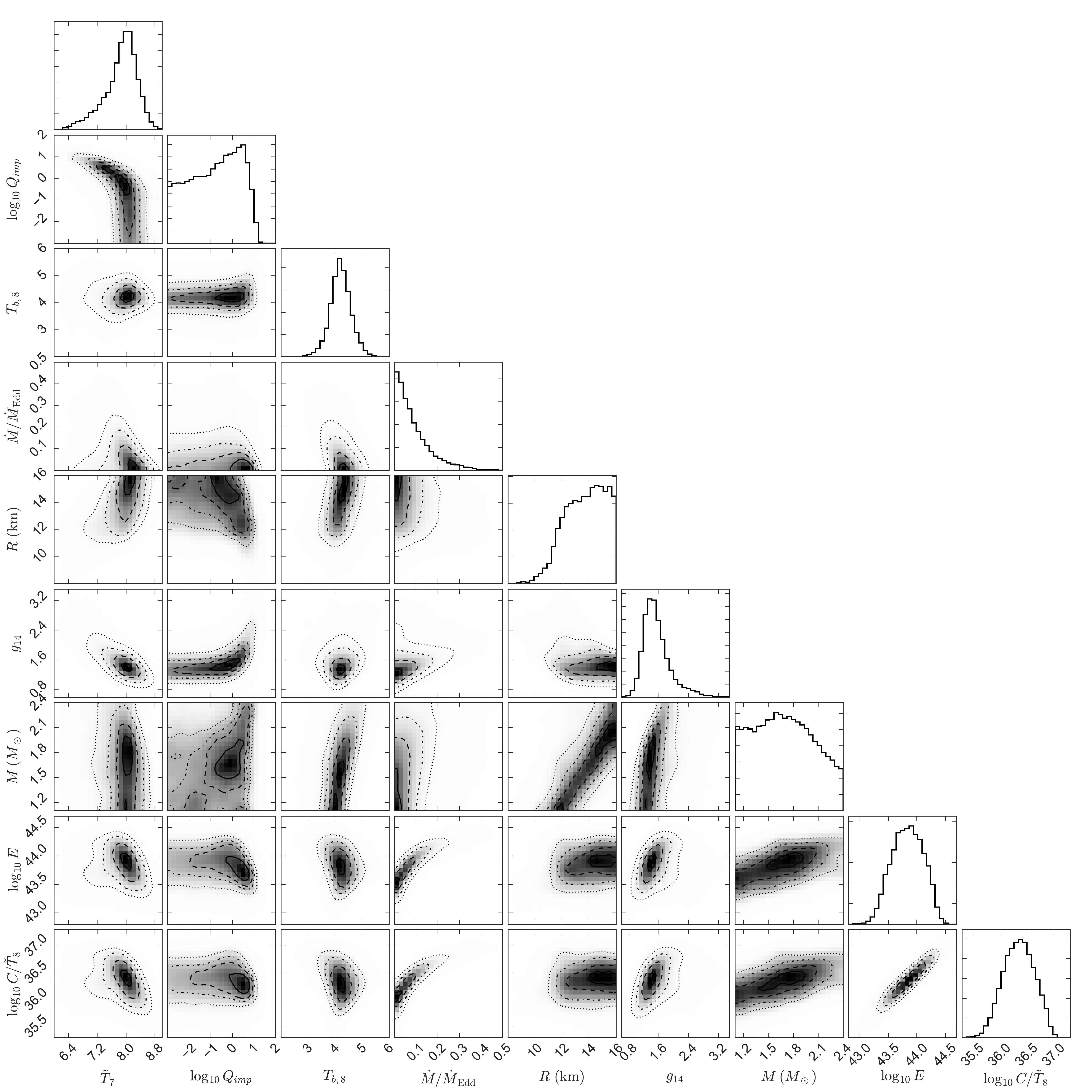} 
\caption{Results of a MCMC fit to the data for \src\ with parameters $\Tc$, $\Qimp$, $T_b$, $\dot M$, $R$ and $g$. Also shown are the resulting distributions of $M$, $E$ and the lower limit on core heat capacity $C$. The contours show where the probability falls to $\exp(-x^2/2)$ of its peak value, for $x=0.5, 1, 1.5, 2$.
\label{fig:mcplot}}
\end{figure*}

Figure \ref{fig:mcplot} shows the posterior distribution of the parameters $\Tc$, $T_b$, $\Qimp$, $\dot M$, $R$ and $g$, as well as the corresponding values of $M$, $E$ and $C$. We see the same upper limit on the impurity parameter $\Qimp\lesssim 10$ found in \cite{Brown2009}. 
Low values of accretion rate $\dot M\lesssim 0.4\ \dot M_{\rm Edd}$ are preferred, and as found by \cite{Brown2009}, values of $\dot M=0$ with no deep heating are allowed, in which case the crust is heated entirely from above by the heat flux entering at the upper boundary. The preferred values of $M$ and $R$ are similar to those in Ref.~\cite{Brown2009} (see Fig.~14 of that paper), with $R\approx 13\ {\rm km}$ for a $1.4\ M_\odot$ neutron star. Taking the mean and standard deviation of the distribution of $C$, we find $\log_{10}\ (C/ \Tc_8\ {\rm erg\ K^{-1}})= 36.4\pm 0.3$. The distribution peaks at $C\approx 2.5\times 10^{36}\ {\rm erg\ K^{-1}}\Tc_8$, similar to Eq.~(\ref{eq:Climit}), but values $C\approx 10^{36}\ {\rm erg\ K^{-1}}\Tc_8$ and as large as $C\approx 10^{37}\ {\rm erg\ K^{-1}}\Tc_8$ are allowed for some parameter ranges. 

\subsection{Effect of accretion rate variations during outburst}
\label{s.accretion-rate-variations}

The quiescent cooling curve depends on the temperature profile in the crust at the end of the outburst. However, as emphasized by \cite{Ootes2016}, the accretion rate during the outburst was variable. The observed long term light curve plotted in Figure 1 of \cite{Cackett2006} shows that the luminosity in the last two years of the outburst was about three times smaller than at the peak. In the simulations so far we took the accretion rate to be constant at the mean value during the outburst. For the deep heating in the inner crust where the thermal time is long, this is a good approximation, but the shallow heat source can respond quickly (thermal timescale of tens of days) to accretion rate variations. Ref.~\cite{Ootes2016} showed that neglecting accretion rate variations results in a factor of two underestimate of the shallow heating strength (see also \cite{Deibel2015} for another example of this effect). Since $E$ is dominated by deep heating, we might expect that it would be less sensitive to accretion rate variations. 

To investigate the effect of accretion rate variations on $E$, we ran a model in which the accretion rate during the last two years was half the average value (with the accretion rate during the first ten years adjusted to maintain the same average over the outburst). Instead of holding the temperature at the top of the crust fixed, we included an extra shallow heat source at the same density $\rho\approx 4\times 10^8\ {\rm g\ cm^{-3}}$ as \cite{Ootes2016}, finding similar values for the shallow heat strength needed to match the cooling curve (1.3 MeV with accretion rate variations, 0.8 MeV without). The dashed curve in Figure \ref{fig:Lin} shows the luminosity entering the core as a function of time. It rises to a larger luminosity initially because of the 20\% larger accretion rate than in the constant accretion rate case. After 10 years, the accretion rate drops and the luminosity evolves towards a new equilibrium, falling below the constant accretion rate curve. In total, we find $E=8.5\times 10^{43}\ {\rm erg}$, as compared to $7.5\times 10^{43}\ {\rm erg}$ previously, so that while the shallow heat source varies by almost a factor of two between the two cases, the energy entering the core changes by only about 10\%. Accretion rate variations are important for correctly deducing the size of the shallow heat source, but do not significantly change the amount of energy deposited into the core. 

\section{Comparison with predictions for the heat capacity of a neutron star core}\label{s.comparison-with-predictions}

We now compare our derived lower limit on the core heat capacity with expected values from equation of state models. We start by estimating the expected magnitude of $C$ from different particle species in the core (Sec.~\ref{s.expected-size-C}), and then calculate the heat capacity of detailed neutron star models for a specific equation of state for different masses and including a density-dependent superfluid gap (Sec.~\ref{s.detailed-calc-C}). We compare to observations in Sec.~\ref{s.comparison-observations}.

\subsection{Expected size of heat capacity}\label{s.expected-size-C}

In dense matter at low temperature heat is primarily carried by fermionic particle-hole excitations at the Fermi surface, providing a heat capacity per unit volume $\sum_i c_{F,i}$, the sum being over fermion species $i$,
with
\begin{eqnarray}\label{eq:cFi}
c_{F,i} = N_i(0) \frac{\pi^2}{3} k_{B}^2 T = \frac{m_i^* p_{\mathrm{F},i}}{3 \hbar^3} k_B^2 T \simeq
\nonumber \\
2\times 10^{19} \left(\frac{m_i^*}{m_n}\right) \left(\frac{p_{\mathrm{F},i}}{400~\rm MeV}\right) 
T_8  \frac{\rm ergs}{\rm  cm^3~K}
\label{eq:CvF2}
\end{eqnarray}
where $p_{\mathrm{F},i}$, $m_i^*$, and $N_i(0)$, are the Fermi momentum, Landau effective mass and density of state at the Fermi surface, respectively, of species $i$, and $T_8$ is the local temperature in units of $10^8\ {\rm K}$.
Phase transitions to a superfluid state can give rise to bosonic collective excitations (Goldstone bosons), but these are much smaller than the fermionic contribution. (The specific heat per unit volume from these excitations at low temperature is given by  
\begin{equation}
c_{B,i}=\frac{2\pi^2~c^3}{15~v^3_{B,i}} ~(k_B T)^3 \simeq 1.5\times 10^{10}  \left(\frac{c}{v_{B,i}}\right)^3~T^3_8~ \frac{\rm ergs}{\rm  cm^3~K}.
\label{eq:CvB}
\end{equation}
where $v_{B,i}$ is the velocity of the Goldstone Boson with linear dispersion relation $\omega =  v_{B,i} k$, and $c$ is the speed of light.)     

The neutron star's total heat capacity is  
\begin{equation}\label{eq:Cintegral}
C = \int_0^R \frac{4\pi r^2 \sum_i c_{i}
dr}{\sqrt{1-2G M(r)/(c^2r)}} \;.
\end{equation}
We can write this as $C=k(4\pi R^3/3)\sum_i c_{i,c}$, where $c_{i,c}$ is the heat capacity of species $i$ at the center of the star and $k$ is an order unity constant (since the density is slowly varying with radius through much of the core). For example, assuming a density profile of the form $\rho(r)=\rho_c\left[1-(r/R)^2\right]$ (the Tolman VII equation of state \cite{Tolman1939}), assuming $c_i\propto \rho^{1/3}$, and neglecting the general relativistic factor in Eq.~(\ref{eq:Cintegral}) gives $k=0.688$ \cite{Yakovlev2003}.
Using this value and considering the neutrons only, since they dominate by number, with the effective mass of the neutron assumed to be $m_N^*=0.7m_N$, we find 
\begin{eqnarray}\label{eq:Canal}
C&=&1.3\times 10^{38}\ {\rm erg\ K^{-1}}\ \Tc_8\ f_c\ \times\nonumber\\&& \left(\frac{R}{12\ {\rm km}}\right)^2\left(\frac{M}{1.4\ M_\odot}\right)^{1/3}\left({1+z\over 1.24}\right)^2.
\end{eqnarray}
In calculating Eq.~(\ref{eq:Canal}), we include two global redshift factors $(1+z)^2=(1-2GM/Rc^2)^{-1}$ to approximate the change in the volume element in equation (\ref{eq:Cintegral}) and to convert the redshifted core temperature to the local temperature at the center of the star; in addition, the scaling $C \propto \mathrm{Volume} \, \cdot \rho_c^{1/3}$ is written as $R^2 M$.
Following \cite{Yakovlev2003}, we include the scaling factor $f_c$ to account for other species. Adding the proton contribution typically gives $f_c\approx 1.25$.

It is well known the pairing between nucleons at the Fermi surface is possible and leads to p-wave  neutron superfluidity, and s-wave proton superconductivity in the core, and will result in an exponential suppression of their respective contributions to the heat capacity 
\citep[for a review, see][]{page2014}. 
The suppression $\propto \exp{(-2\Delta/T)}$, where $\Delta \simeq 1 \MeV$ is the superfluid/superconducting pairing gap, is severe at low temperature and can suppress the nucleon contribution to well below that expected for leptons. 
This implies that when the NS core is composed entirely of nucleons and leptons, the lepton contribution provides a robust lower limit on the total heat capacity, corresponding to $f_c\sim 0.1$ in Eq.~(\ref{eq:Canal}).

In a larger class of models that include phase transitions to matter containing quarks, hyperons, or meson condensates, the population of leptons necessary for electric neutrality is reduced by the additional negatively charged hadronic or quark components. This reduction is modest in most models and we conclude that the lower limit on the heat capacity, set by the lepton contribution, is still $\gtrsim 10^{37}\ {\rm ergs/K}\ \Tc_8$. 
There is, however, one interesting exception. The color-flavor-locked (CFL) phase of dense quark matter contains equal numbers of up, down, and strange quarks and thereby does not need leptons in the ground state to ensure electric charge neutrality (see Ref.~\cite{Alford:2007xm} for a comprehensive review of color superconducting phases of dense quark matter). This phase is a color superconductor, and all nine quark (3 flavors $\times$ 3 colors) participate in BCS pairing due to an attractive interactions mediated by the exchange of gluons. The preferred pairing pattern in which quarks of different flavors pair ensures equal number densities for all flavors of quarks. Further, since all of the quarks are paired, their contribution is exponentially suppressed by the factor $\propto \exp{(-2\Delta/T)}$ where $\Delta\simeq 10\textrm{--}100 ~\MeV \gg T $ is the pairing gap in dense quark matter, and the heat is carried by Goldstone bosons. These Goldstone bosons arise naturally due to the breaking of global symmetries in the superconducting state, and their velocities $v_B\approx c/\sqrt{3}$ are large. From Eq.~(\ref{eq:CvB}) we can deduce that this contribution to the specific heat of the CFL phase will be about eight orders of magnitude smaller than the electron contribution in nuclear matter at $T=10^8$ K. Consequently, were CFL quark matter to occur in the inner parts of the NS core their contribution can be discounted. In models where CFL quark matter occupies a large fraction of the core the total heat capacity of the star would be well below the usual lepton bound, $C\ll 10^{37}\ {\rm erg/K} \, \Tc_8$.

\subsection{Detailed calculation of core heat capacity}\label{s.detailed-calc-C}

To properly calculate the heat capacity of the neutron-star core from a particular nuclear interaction self-consistently for different masses, and to include the density-dependence of the superfluid gaps, we have employed the IU-FSU relativistic mean-field model \cite{Fattoyev:2010mx}. In the RMF model the nucleon effective mass (aka Dirac mass) then becomes a function of the mean-field baryon density, $m_{\rm D}^{*} = m + \Sigma_{\rm s}(\rho)$, where $\Sigma_{\rm s}(\rho)$ is the nucleon scalar self-energy. The Landau effective mass at a fixed baryon density is given through $m^* = \sqrt{m_{\rm D}^{*2} + p_{\rm F}^2/c^2}$. The employed IU-FSU parameterization was originally derived with the goal of softening the symmetry energy to generate smaller neutron star radii, and stiffening the overall EOS at higher density to generate a larger limiting neutron-star mass. This model predicts a relatively small neutron-skin thickness of $\mathrm{^{208}Pb}$ (see Table~\ref{Table1})---a fundamental nuclear-structure observable that will be measured with increasing accuracy at the Jefferson Laboratory~\cite{PREXII:2012}.  Although the extensive experimental database of nuclear masses and charge radii is sufficient to constrain most of the bulk parameters of neutron-rich
matter, it is insufficient to constrain those associated with the density dependence of the symmetry energy.

By tuning two purely isovector parameters of the RMF model \cite{Horowitz:2000xj, Fattoyev:2012ch} one can generate a family of model interactions that are almost indistinguishable in their predictions for a large set of the nuclear ground state observables that are mostly isoscalar in nature, yet predict different isovector observables such as the neutron skin of neutron-rich heavy nuclei. As a contrast to the original IU-FSU model we use two additional interactions with varying density dependence of the symmetry energy (see Table~\ref{Table1}).  \citet{Piekarewicz:2014lba} showed that models which predict intermediate values of the slope of the symmetry energy predict large values of the crust-core transition pressure; as a result, the neutron star crust is thicker and the crustal fraction of the moment of inertia is larger.  Thicker crusts are consistent with the observation of pulsar glitches, when crustal entrainment effects are taken into account~\cite{Andersson:2012iu, Piekarewicz:2014lba}. Owing to its maximum transition pressure of $P_{\rm t} = 0.518\,\mathrm{MeV\,fm^{-3}}$ we refer to this model as IU-FSU (max). We should mention that both the original IU-FSU model and the IU-FSU (stiff), which has a stiff slope
of the symmetry energy ($L = 95\,\mathrm{MeV}$), predict almost twice smaller crust-core transition pressures $P_{\rm t} = 0.289\,\mathrm{MeV\,fm^{-3}}$ and $P_{\rm t} = 0.293\,\mathrm{MeV\,fm^{-3}}$,
respectively. The corresponding transition densities for these models are anticorrelated with the value of the slope of the nuclear symmetry energy at saturation, $\rho_{\rm t} = 0.087\mathrm{fm^{-3}}$ for IU-FSU, $\rho_{\rm t} = 0.077\,\mathrm{fm^{-3}}$ for IU-FSU (max), and $\rho_{\rm t} = 0.057\,\mathrm{fm^{-3}}$ for IU-FSU (stiff).

\begin{table}
\renewcommand{\arraystretch}{0.9}
\begin{tabular}{lrrrrrrl}
 \hline\hline
 \rule{0pt}{2.6ex}Model & \tabhead{$J$} & \tabhead{$L$} & \tabhead{$K_{\rm sym}$} & \tabhead{$B/A$} & \tabhead{$R_{\rm ch}$} & \multicolumn{2}{c}{$R_{\rm skin}$}  \\
 & \tabhead{[MeV]} & \tabhead{[MeV]} & \tabhead{[MeV]} & \tabhead{[MeV]} & \tabhead{[fm]} & \multicolumn{2}{c}{[fm]} \\
\hline 
\rule{0pt}{2.6ex}IU-FSU    & 31.30 & 47.20
               & $+$28.53 & $-$7.89 & 5.49 & 0.16 & \\
IU-FSU (max)   & 33.88 & 65.00
               & $-$60.30 & $-$7.89 & 5.47 & 0.22 & \\
IU-FSU (stiff)   & 37.02 & 95.00
              & $-$57.86 & $-$7.87 & 5.46 & 0.27 & \\
\hline
\rule{0pt}{2.6ex}Experiment   &  &  &  & $-$7.87 & 5.50 & $0.33$ & $\!\!{}^{+0.16}_{-0.18}$ \\[0.2em]
\hline\hline
\end{tabular}
\caption{Predictions for the bulk parameters characterizing the
behavior of infinite nuclear matter at saturation density
$\rho_{_{0}} = 0.1546\,\mathrm{fm^{-3}}$. The binding energy per nucleon and incompressibility coefficient of symmetric nuclear matter are identical in these interactions with $\varepsilon_{_{0}} = -16.40\,\MeV$ and $K_{0} = 231.33\,\mathrm{MeV}$, respectively, whereas $J$, $L$, and $K_{\rm sym}$ which represent the energy, slope, and curvature of the symmetry energy (see \cite{Horowitz2014} for definitions), are quite different. Also shown are the binding energy per nucleon $B/A$, charge radius $R_{\mathrm{ch}}$, and neutron-skin thickness $R_{\mathrm{skin}}$ of $\mathrm{^{208}Pb}$, along with their corresponding experimental values.
}
\label{Table1}
\end{table}

In Table~\ref{Table2} predictions for the total and core heat capacity for various mass neutron stars are given. In calculating these numbers we assumed an isothermal star with redshifted constant temperature of $\Tc = 10^8\,\mathrm{K}$. 
Notice that the core heat capacity scales linearly with the temperature, whereas there is a slight deviation from linearity between the total heat capacity and the temperature due to the contribution from ions in the crust. The total heat capacity is given assuming that the neutrons in the crust are normal; in practice, they are expected to be superfluid, which reduces the crust heat capacity by about an order of magnitude compared to the value in Table~\ref{Table2}.

If all nucleons in the core are in a superfluid state, then the core heat capacity contains only contributions from leptons and is drastically reduced. As an example, for a $1.4\,M_\odot$ neutron star in the IU-FSU model superfluidity reduces the heat capacity from $C_{\rm core} = 1.709 \times 10^{38}\,\mathrm{erg\,K^{-1}}$ to $C_{\rm core} = 0.194 \times 10^{38}\, \mathrm{erg\,K^{-1}}$, which is almost an order of magnitude lower (see Table~\ref{Table2}). 

\begin{table}
\renewcommand{\arraystretch}{0.9}
\begin{tabular}{rlrrrrrrrr}
  \hline\hline
 \tabhead{$M$}  & \rule{0pt}{2.6ex}Model & \tabhead{$R_{\rm tot}$} & \tabhead{$R_{\rm core}$} & \tabhead{$\rho_c$} & \tabhead{$C_{\rm tot}$} & \tabhead{$C_{\rm core}$}  & \tabhead{$C_{\rm core, l}$}   \\
 \tabhead{$[M_{\odot}]$} & & \multicolumn{2}{c}{[km]} & \tabhead{[$\mathrm{fm^{-3}}$]} & \multicolumn{3}{c}{[$10^{38}\,\mathrm{erg\,K^{-1}}$]} & \\
  \hline
\rule{0pt}{2.6ex} $1.20$ & APR            & 11.853 & 10.430 & \rule{0.8em}{0ex}0.490 & \rule{0.8em}{0ex}1.379 & 1.368 & 0.139 \\
         & IU-FSU         & 12.519 & 11.283 & 0.404 & 1.583 & 1.469 & 0.152 \\
         & IU-FSU (max)   & 12.895 & 11.379 & 0.402 & 1.698 & 1.517 & 0.165 \\
         & IU-FSU (stiff) & 13.525 & 12.033 & 0.382 & 1.897 & 1.741 & 0.182 \\[0.5em]
  $1.40$ & APR            & 11.708 & 10.586 & 0.544 & 1.596 & 1.586 & 0.178 \\
        & IU-FSU          & 12.511 & 11.504 & 0.471 & 1.813 & 1.709 & 0.194 \\
         & IU-FSU (max)   & 12.804 & 11.581 & 0.470 & 1.923 & 1.761 & 0.212 \\
         & IU-FSU (stiff) & 13.328 & 12.141 & 0.452 & 2.120 & 1.985 & 0.238 \\[0.5em]
  $1.60$ & APR            & 11.560 & 10.669 & 0.607 & 1.830 & 1.821 & 0.226 \\
         & IU-FSU         & 12.406 & 11.592 & 0.567 & 2.048 & 1.956 & 0.244 \\
         & IU-FSU (max)   & 12.629 & 11.649 & 0.569 & 2.153 & 2.011 & 0.267 \\
         & IU-FSU (stiff) & 13.064 & 12.124 & 0.548 & 2.345 & 2.230 & 0.304 \\[0.5em]
  $1.80$ & APR            & 11.371 & 10.681 & 0.683 & 2.099 & 2.091 & 0.287 \\
         & IU-FSU         & 12.109 & 11.476 & 0.743 & 2.291 & 2.213 & 0.308 \\
         & IU-FSU (max)   & 12.264 & 11.509 & 0.747 & 2.388 & 2.270 & 0.336 \\
         & IU-FSU (stiff) & 12.625 & 11.906 & 0.718 & 2.574 & 2.481 & 0.386 \\
\hline\hline
\end{tabular}
\caption{Predictions for the properties of various mass neutrons stars, such as the total radius $R_{\rm tot}$, the core radius $R_{\rm core}$, the central baryon density $\rho_c$, total heat capacity  $C_{\rm tot}$, core heat capacity when nucleons are in the normal state $C_{\rm core}$, and with just lepton contributions only $C_{\rm core, l} = C_{\rm core, e}  + C_{\rm core, \mu}$---corresponding to the case when all nucleons are in the superfluid state. An isothermal star with $\Tc = 10^8\,\mathrm{K}$ is assumed. The heat capacities have contributions from ions, electrons, muons, protons and neutrons, the latter four species being in beta-equilibrium at the core of the neutron star. The total heat capacity assumes that the crust neutrons are normal.}
\label{Table2}
\end{table}

\begin{figure}[htbp]
   \includegraphics[width=\columnwidth]{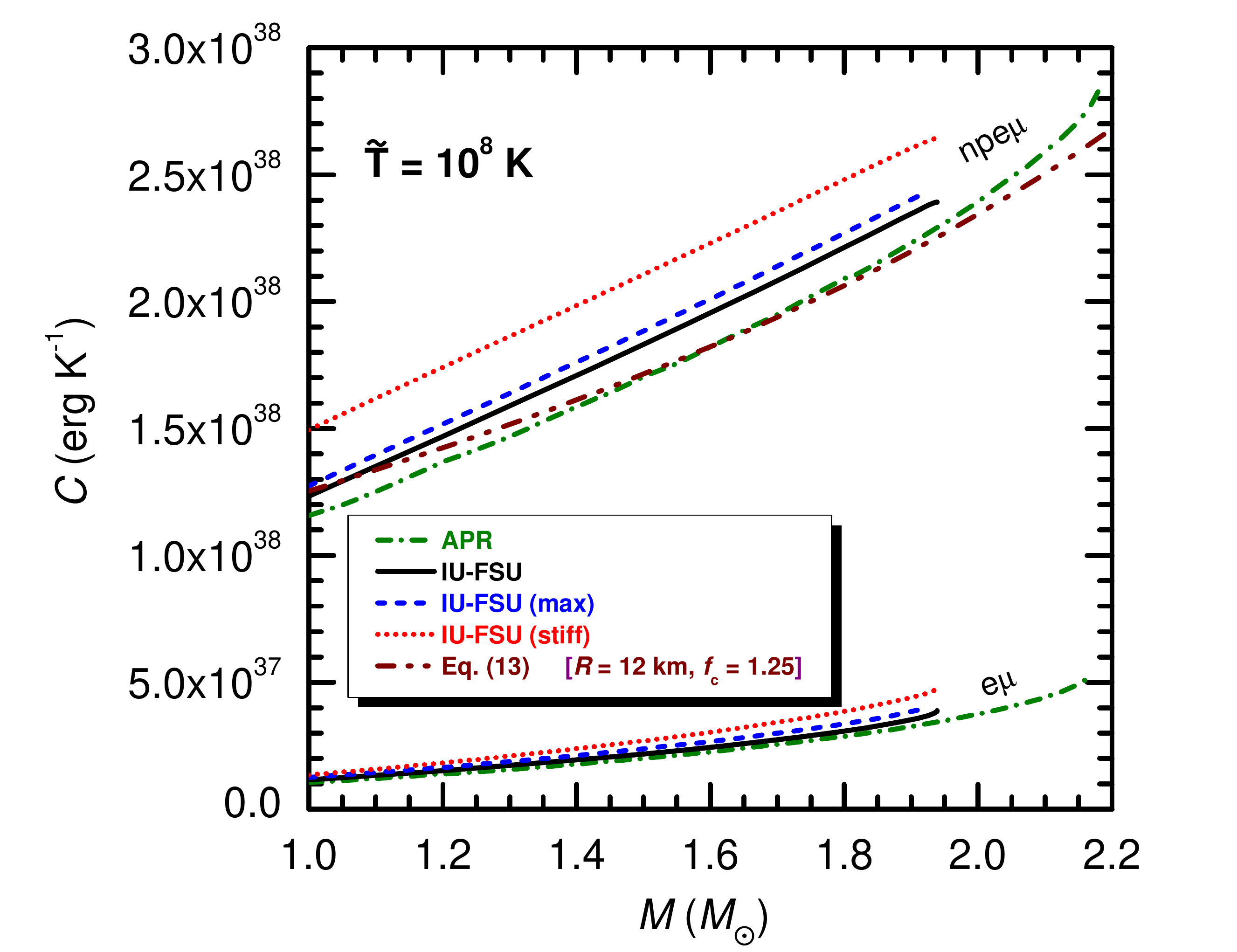} 
   \caption{The core heat capacity $C$ as a function of neutron-star mass $M$ for the APR equation of state and the three RMF models discussed in the text: from top to bottom these are IU-FSU (stiff) with red dotted line; IU-FSU (max) with blue dashed line; IU-FSU with black solid line; and APR with green dash-dotted line. Also is shown (brown dash-dot-dotted line) is the result from Eq.~(\ref{eq:Canal}) with a constant radius $R = 12\,\mathrm{km}$ and scaling factor $f_c = 1.25$. The core heat capacities labeled ``$npe\mu$'' are due to all particle species, whereas those labeled ``$e\mu$'' are due to leptonic contributions only.}
   \label{Fig1CvMass} 
\end{figure}

Figure~\ref{Fig1CvMass} shows the core heat capacity as a function of the neutron-star mass for the three RMF interactions and the APR EOS. The displayed result broadly brackets the range of the core heat capacity due to variations of the EOS. For example, the core heat capacity of a canonical $1.4\,M_\odot$ neutron star is in the range $1.60 \, \Tc_8 < C_{38} < 2.12 \, \Tc_8$, where 
$C_{38} = C/(10^{38}\,\mathrm{erg\,K^{-1}})$. The heat capacity increases roughly linearly with mass, coming from a combination of the scalings with $M^{1/3}$ and redshift in Eq.~(\ref{eq:Canal}).

\begin{figure}[htbp]
   \includegraphics[width=\columnwidth]{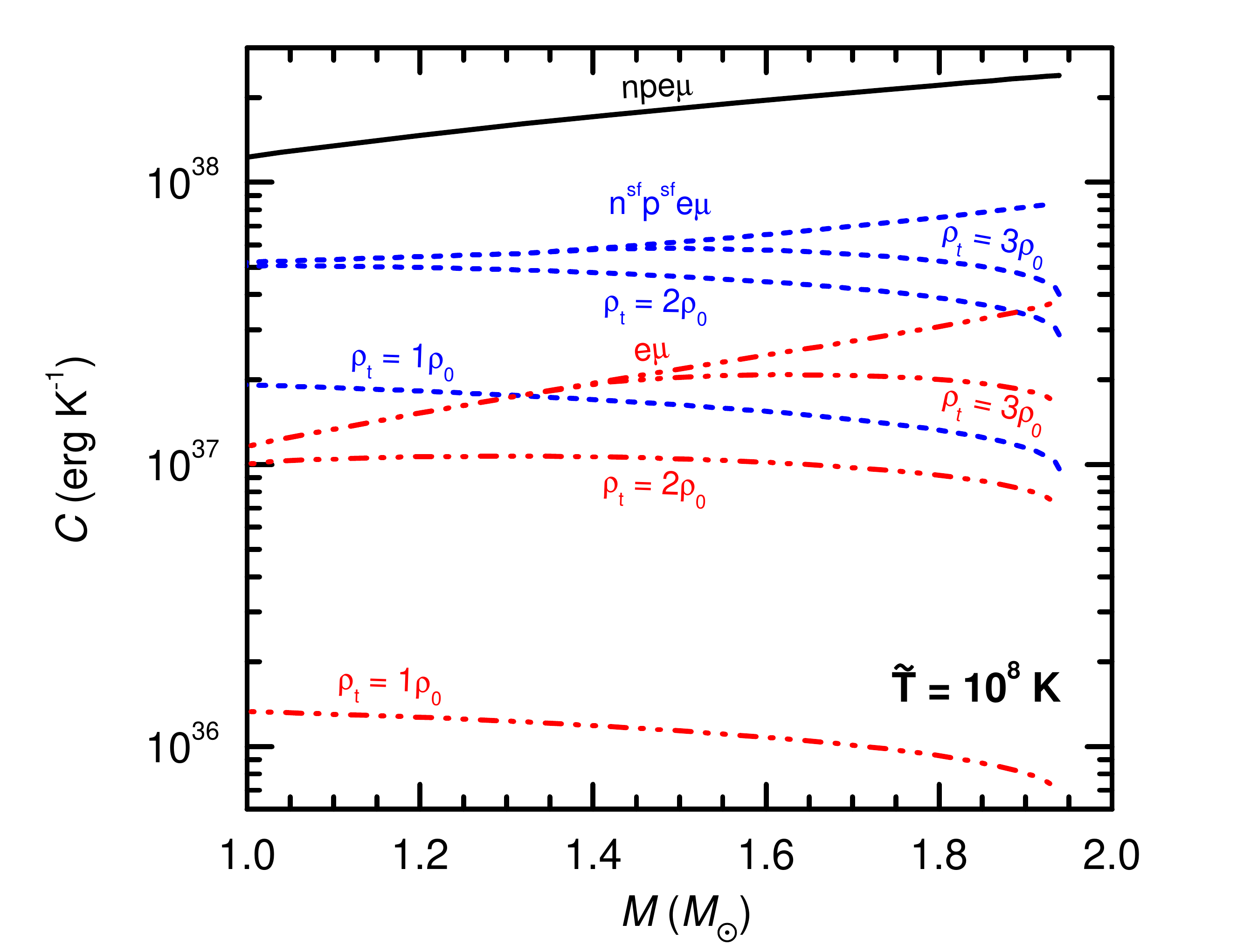} 
   \caption{The core heat capacity as a function of neutron-star mass is shown for various states of neutron-star matter with IU-FSU as reference model. The black solid line corresponds to the $npe\mu$ matter when nucleons are in normal state, the blue short-dashed line corresponds to the case when protons are in $^1S_0$ superconducting state whereas neutrons are in $^3P_2$ superfluid state using the pairing gap model of Ref.~\cite{Baldo:1992}, and the red dash-dot-dotted line corresponds to the case when all neutrons and protons are superfluid in the core. The curves labeled with $\rho_{\rm t}$ show the heat capacity when there is a transition to an exotic state of matter with vanishingly small specific heat, such as the CFL phase of quark matter, at a baryon density $\rho_{\rm t}$.}
   \label{Fig2CvMass} 
\end{figure}

\begin{figure}[htbp]
   \includegraphics[width=\columnwidth]{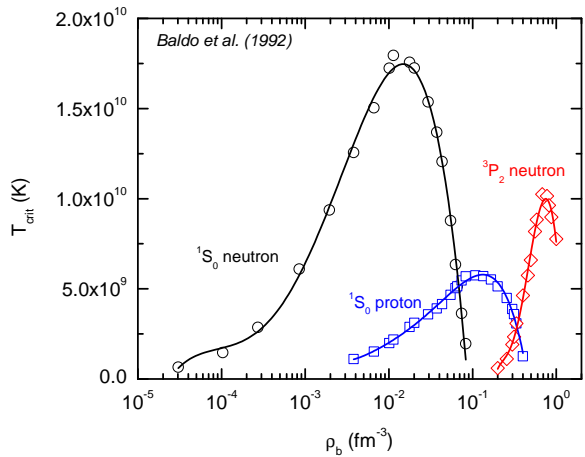} 
   \caption{The critical temperature for the various kinds of superfluidity in the neutron star matter as a function of baryon density. The figure is derived using the pairing gap model of Ref.~\cite{Baldo:1992}. The boundary between the inner crust and the core is located at $\rho_b\approx 0.05$--$0.1\ {\rm fm^{-3}}$.}
   \label{TcritBaldo1992} 
\end{figure}

Taking the IU-FSU as a reference model, we also display in Figure \ref{Fig2CvMass} the core heat capacity as a function of neutron-star mass for various phases of matter in the core. Microscopic theories of dense matter suggest that protons suffer a singlet $^1S_0$ state pairing at intermediate nuclear densities of about $0.5\rho_0$ to about few $\rho_0$ , where $\rho_0$ is nuclear saturation density. On the other hand, neutrons form pairs in the $^1S_0$ state at densities pertaining to the crust of neutron stars. As the density increases the effective neutron-neutron attraction becomes repulsive, and at $\rho \gtrsim \rho_0$ the effective attraction again develops between neutrons as a result of the triplet $^3P_2$ state. 

The density-dependence of the pairing gap, and hence the superfluid critical temperature, is quite uncertain and model dependent \citep{page2014}. 
For example, one interpretation of the possible rapid cooling of the neutron star in Cassiopeia A involves the development of the neutron $^3P_2$ superfluid state in the presence of extensive proton $^1S_0$ superconductivity \citep{page2011}: in such a model, given the low temperature of {\src}'s core, the heat capacity of both neutrons and protons would be strongly suppressed in the whole core and only the lepton contribution would remain.
As a alternative example, consider the density-dependence of the pairing gap from~\cite{Baldo:1992} (see Fig.~\ref{TcritBaldo1992}). A significant fraction of protons and neutrons are then paired in the stellar interior. For $T < T_{\rm crit}$, the nucleon specific heat capacity can be written as $c_{\rm N} = c_{\rm N0} f(T)$, where $c_{\rm N0}$ is the specific heat capacity of normal nucleons, and the factor $f(T)$ describes the variation of heat capacity by superfluidity~\cite{Levenfish:1994}. Following the same method as outlined in Ref.~\cite{Levenfish:1994}, we calculated the heat capacity of the neutron star when both neutrons and protons are superfluid. The core heat capacity is reduced, but remains $\approx 2$--$4$ times larger than the lepton-only value. For a $1.4 \, M_{\odot}$ neutron star, about 3.6\% of protons and 33.4\% of neutrons remain normal due to the density dependence of the pairing gap (see Fig.~\ref{TcritBaldo1992}).
This shows that depending on the gap model, the heat capacity may not be as small as the lepton-only value, which corresponds to the extreme case of superfluidity, when all nucleons are superfluid in the entire region of the star.
Including a transition to a CFL phase at a density $\rho_t$, where we assume that the CFL phase does not contribute to the heat capacity, can significantly reduce the heat capacity if $\rho_t\lesssim 2\ \rho_0$.


\subsection{Comparison with observations}\label{s.comparison-observations}

\begin{figure}[htbp]
\includegraphics[width=\columnwidth]{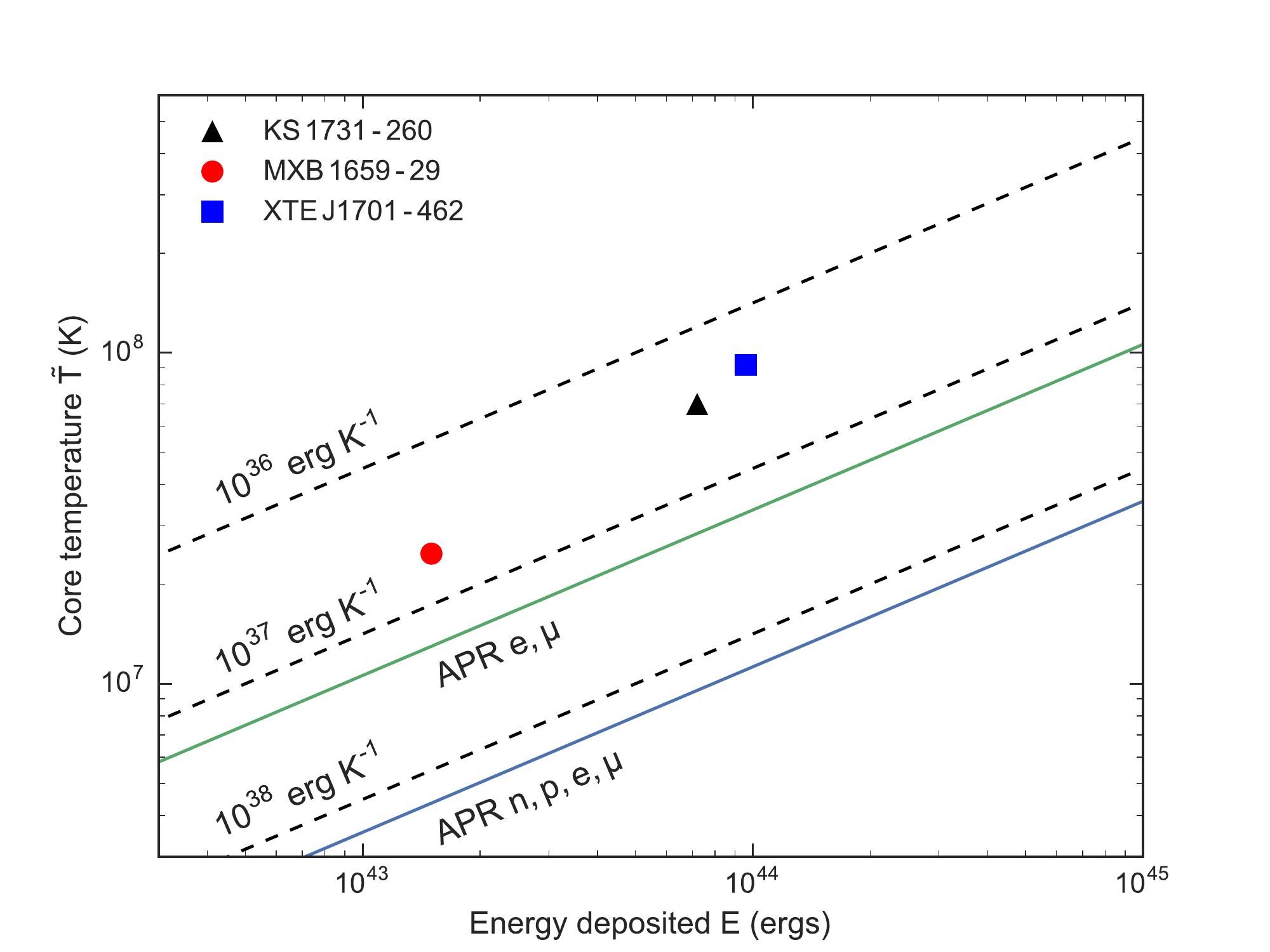} 
\caption{The core temperature reached by an initially cold core as a function of the energy deposited. The data points are the derived energy and core temperature for \src, \mxb, and \xtej. For \src\ we assume a Fe envelope, and for \mxb\ and \xtej\ we assume a He envelope (as indicated by fits to their cooling curves). The green and blue solid lines are for the heat capacity of a $1.4 M_\odot$ and APR equation of state: the upper curve is for electrons and muons only, the lower curve includes electrons, muons, protons and neutrons. The dashed lines show values of $C=10^{36}$, $10^{37}$ and $10^{38}\ \Tc_8\  {\rm erg\ K^{-1}}$.\label{fig:core1}}
\end{figure}

\begin{figure}[htbp]
\includegraphics[width=\columnwidth]{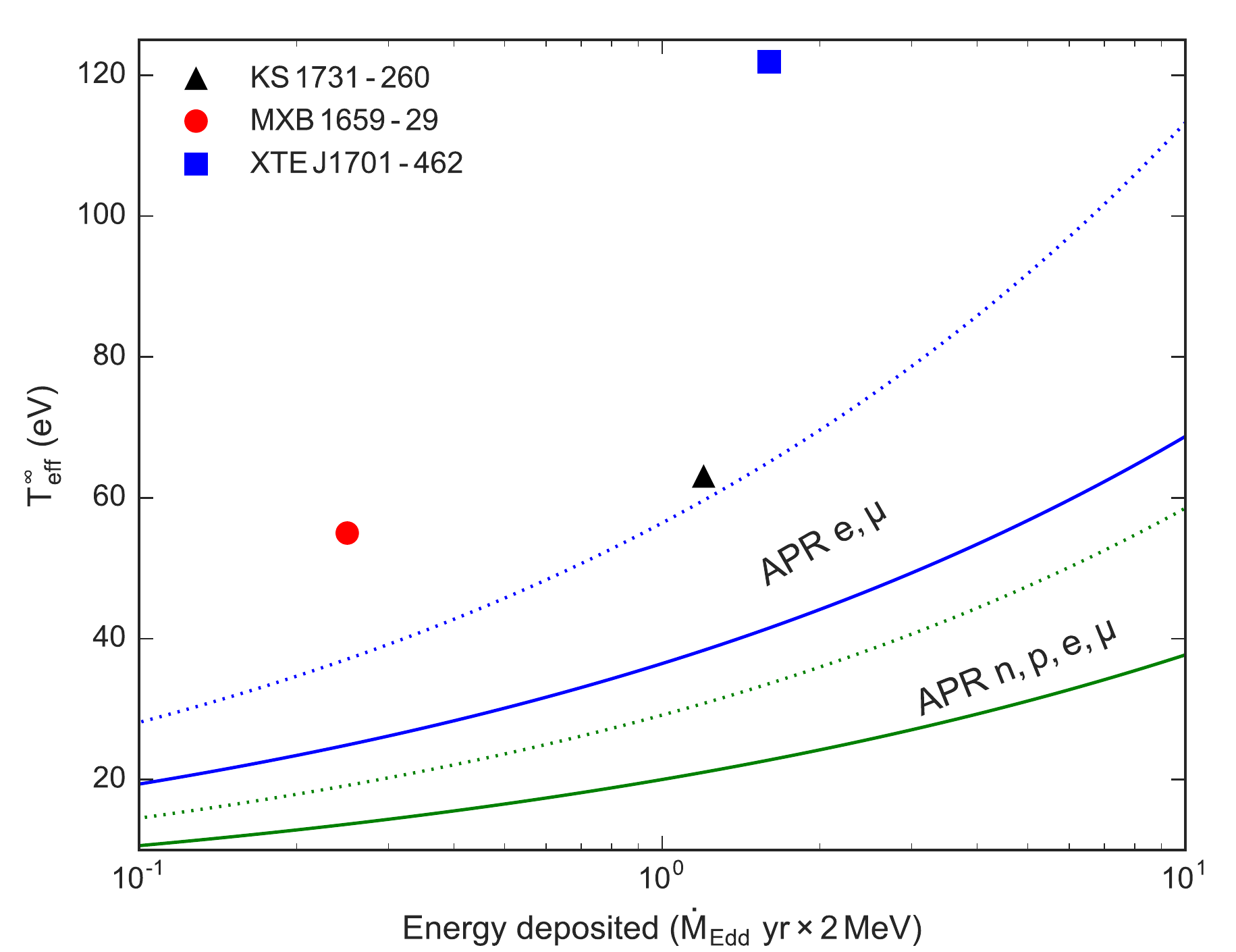} 
\caption{Observed quiescent temperature vs.\ the deposited energy for an outburst with $\dot M = \dot M_{\rm Edd}$ for 1 year and $\Qnuc = 2\ {\rm MeV}$ per nucleon. The curves are for APR $1.4 M_\odot$ with and without nucleon pairing as in Fig.~\ref{fig:core1}. In each case, the dotted curve assumes a light element envelope, and the solid curve a heavy element envelope.
\label{fig:core2}}
\end{figure}

Figs.~\ref{fig:core1} and \ref{fig:core2} show two different ways to compare these theoretical expectations with observations. For the observations, we include the lower limit on $C$ for \src\ derived in Sec.~II, and also add two other sources \mxb\ and \xtej. \mxb\ has the lowest temperature of observed quiescent transients, while \xtej\ had a larger accretion rate and shorter outburst by about an order of magnitude compared to \src\ and \mxb. Given the good agreement with the value of $E$ we inferred from our lightcurve modeling for \src\ and the value estimated from equation (\ref{eq:E}), we use equation (\ref{eq:E}) to estimate $E$ for the other two sources. Other sources have observed cooling curves, but are not as constraining and so we do not include them. These include MAXI~J0556-332 which had a high accretion short outburst similar to \xtej, but with a much larger temperatures \cite{Homan2014}, and EXO~0748-676 with a long but weaker outburst \cite{Degenaar2014}.

The outburst properties are given in Table \ref{tab:obs}, together with the values of $\Tc$, $E$, and $C$ that we infer using Eq.~(\ref{eq:E}) for $E$, Eqs.~(\ref{eq:Tc_heavy}) or (\ref{eq:Tc_light}) for $\Tc$, and Eq.~(\ref{eq:Climit}) for $C$. For \src, we assume an Fe envelope as discussed in \S II, but for the other two sources take a light element envelope that has been successfully used to fit their cooling curves (see~\cite{Brown2009} for \mxb\ and \cite{Page2013} for \xtej). For \mxb, we take $\Teffinf=55\ {\rm eV}$, which is the value obtained in Ref.~\cite{Cackett2013} by allowing the X-ray absorption column $N_H$ to be free in the fit to the observed spectrum. If instead $N_H$ is held fixed, the temperature is lower $\Teffinf = 49\ {\rm eV}$ \cite{Cackett2013}, giving a value of $\Tc$ that is 20\% smaller, and $C>5.8\times 10^{36}\ {\rm erg\ K^{-1}}\Tc_8$. However, given the uncertainty in the interpretation of the observed spectrum \cite{Cackett2013}, we choose the more conservative limit.

\begin{table*}
\renewcommand{\arraystretch}{1}
\begin{tabular}{ld{1.1}d{2.1}d{1.1}d{3.1}d{1.2}cd{1.1}c}
 \hline\hline
 \rule{0pt}{2.6ex}Source & \tabhead{$\dot M$} & \tabhead{$\tout$} & \tabhead{$E_{43}$} &
 \tabhead{$\Teffinf$} & \tabhead{$\Tc_8$} & Envelope &  \tabhead{$C$} & Ref. \\
 & \tabhead{$[10^{18}\ {\rm g\ s^{-1}}]$} & \tabhead{$[\mathrm{yr}]$} && \tabhead{$[{\rm eV}]$} & &  composition & \tabhead{$[\Tc_8 10^{36}\ {\rm erg\ K^{-1}}]$} &  \\
\hline
\rule{0pt}{2.6ex}KS~1731-260    & 0.1 & 12 & 7.2 & 63.1  & 0.7 & Fe &  2.9 & \cite{Cackett2010} \\
MXB~1659-29 & 0.1 & 2.5 & 1.5 & 55  & 0.25& He &  4.8 & \cite{Cackett2013}\\
XTE~J1701-462 & 1 & 1.6 & 9.6 & 121.9 & 0.92 & He &  2.2 & \cite{Fridriksson2011} \\
\hline\hline
\end{tabular}
\caption{Observed quiescent temperatures and outburst properties of accreting neutron stars. The energy deposited in the core during outburst $E_{43}$ is determined from Eq.~(\ref{eq:E}), and core temperature $\Tc$ from Eq.~(\ref{eq:Tc_heavy}) or (\ref{eq:Tc_light}) depending on the indicated envelope composition. The lower limit on the heat capacity (normalized to $10^8\ {\rm K}$ assuming $C\propto T$) is from Eq.~(\ref{eq:Climit}).} 
\label{tab:obs}
\end{table*}

Figure \ref{fig:core1} shows the expected $\Tc$ as a function of the energy deposited for different values of $C$. As a specific example, we show the expected temperature if the heat capacity was the value expected for a $1.4\ M_\odot$ star calculated using the APR equation of state with either all particles contributing (neutrons, protons, electrons and muons), or for electrons and muons only \cite{Page1994}. Figure \ref{fig:core2} shows a comparison directly to observed quantities: the expected $\Teffinf$ in quiescence as a function of the outburst properties. For each value of $C$, the two curves show the values of $\Teffinf$ predicted for different envelope compositions (with lighter envelopes having a larger $\Teffinf$ at a fixed core temperature or energy deposited).

The three sources \src, \mxb, and \xtej\ have similar lower limits $C\approx 2$--$5\times 10^{36}\ \Tc_8\ {\rm erg\ K^{-1}}$ (Table \ref{tab:obs}). The most constraining source is \mxb. Despite having less energy deposited than \src\ because of its shorter outburst, the lower observed temperature and helium envelope result in a larger lower limit by almost a factor of two. The lower limits are all consistent with the expected lepton contribution to the heat capacity, but rule out a transition to a CFL phase at densities $\approx 1$--$2\ \rho_0$.

\section{Thermal evolution of the core in outburst and quiescence}\label{s.core-evolution}

The lower limit of the core's heat capacity, Eq.~(\ref{eq:Clim_intro}), implicitly assumes that neutrino emission from the core is negligible.
In this section, we critically examine this assumption and define joint constraints on $C$ and $\Lnu$.  We then show how observational limits on the core cooling during quiescence, perhaps coupled with a measurement of the recurrence time, will further constrain $C$ and $\Lnu$.

\begin{figure}[htbp]
\includegraphics[width=\columnwidth]{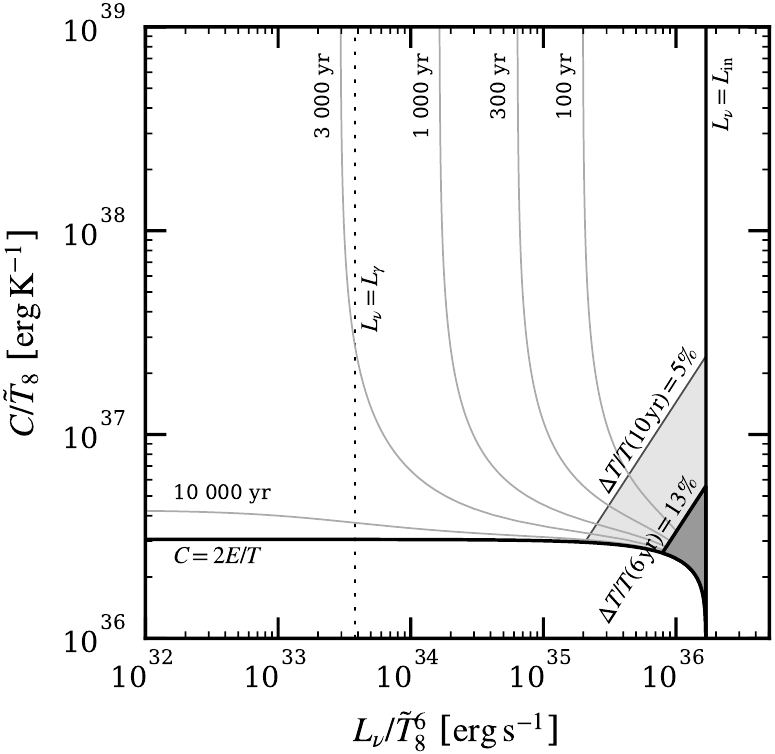} 
\caption{Possible values of the specific heat $C$ and neutrino luminosity $\Lnu$ (assumed $\propto \Tc^6$) for \src.  Neutrino cooling from the core exceeds radiative cooling from the surface to the right of the vertical dotted line. The minimum specific heat is indicated by the lower dark curve; it asymptotically approaches the value derived in Eq.~(\protect\ref{eq:Clim_intro}) for sufficiently small $\Lnu$.  At right, the vertical dark line indicates where $\Lnu(\Tc_8=7)=\Lin$; this is the largest neutrino luminosity compatible with the observed $\Tc$.  The thin grey contours indicate values of constant recurrence time.  The dark grey region at lower right is excluded by the absence of cooling (at $<13\%$) after 6 years in quiescence.  If cooling were absent ($<5\%$) after 10 years, then the light grey region would be further excluded.
\label{fig:phaseplot}}
\end{figure}

The results of this section are presented in Fig.~\ref{fig:phaseplot} for \src. The following subsections describe in more detail the different components of this plot. In brief, the specific heat must lie above the lower dark curve, which approaches the minimum value of Eq.~(\ref{eq:Clim_intro}) for negligible $L_\nu$; the neutrino luminosity must lie to the left of the vertical dark line, which indicates where the core temperature saturates during outburst (Sec.~\ref{s.neutrino-cooling-in-outburst}); the shaded regions indicate constraints from limits on the changes in the core temperature during quiescence (Sec.~\ref{sec:timeevolution}); and the light contours indicate different values of the recurrence time (Sec.~\ref{s.repeated-outbutsts}).

In this section, we make repeated use of the thermal evolutionary equation for the core (Eq.~\ref{eq:evolution}),
\begin{equation}\label{eq:evolution-full}
C\frac{d\Tc}{dt} = -\Lph(\Tc) - \Lnu(\Tc) + \Lin,
\end{equation}
where $\Lin=0$ during quiescence. The photon luminosity, $\Lph(\Tc)$, follows from Equations~(\ref{eq:Tc_heavy}) and (\ref{eq:Tc_light}):
\begin{eqnarray}
\label{e.Lph-heavy}
\Lph &=& 9.8\times 10^{32} \; \Tc_8^{2.2}\,\ergpersec\quad\textrm{(heavy)};\\
\label{e.Lph-light}
\Lph &=& 7.5\times 10^{33}\; \Tc_8^{2.4}\,\ergpersec\quad\textrm{(light)}.
\end{eqnarray}
This equation assumes that the core is isothermal, which holds only if the thermal conduction time across the core is much shorter than the cooling or heating timescale, and the core conductivity is large enough to transport heat inwards with a small temperature contrast. The conduction time across the core is
\begin{eqnarray}
\frac{c_P}{K} R^2 &\sim &  3\ {\rm yr}\ \left({c_P\over 10^{19}\ {\rm erg\ cm^{-3}\ K^{-1}}}\right)\left({R\over 10\ {\rm km}}\right)^2\nonumber\\&&\left({K\over 10^{23}\ {\rm erg\ cm^{-1}\ s^{-1}\ K^{-1}}}\right)^{-1},
\end{eqnarray}
where we insert a typical value of thermal conductivity $K$ due to neutrons at $10^8\ {\rm K}$ \cite{Baiko2001,Shternin2007b} and use the heat capacity of degenerate fermions from Eq.~(\ref{eq:cFi}). This conduction time is a factor of a few times smaller than both the outburst timescale and time in quiescence for \src, and in the case of rapid core evolution with a small $C$, the thermal time is even shorter. The temperature contrast required to transport the inwards luminosity is also small, $\Delta T\approx L/4\pi R K\sim 10^6\ {\rm K}$ for $\Lin \sim 10^{35}\ \ergpersec$, so the isothermal assumption is reasonable.

\subsection{Neutrino cooling during outburst and an upper limit on the core neutrino luminosity}\label{s.neutrino-cooling-in-outburst}

The neutrino emissivity of the neutron star core is highly uncertain, depending on the particle content and allowed weak reactions. A large enough neutrino emissivity would remove a significant amount of the energy deposited in the core during the outburst and invalidate our assumption that all of the energy that flows into the core from the crust heats the core. 

Neutrino cooling processes generally divide into two classes \citep{Yakovlev2003}: fast, such as direct Urca,
\begin{equation}
\epsdUrca \approx 10^{26}\,\mathrm{erg\,cm^{-3}\,s^{-1}} \left(\frac{T}{10^9\,\mathrm{K}}\right)^6;
\end{equation}
and slow, such as modified Urca,
\begin{equation}
\epsmUrca \approx 10^{20}\,\mathrm{erg\,cm^{-3}\,s^{-1}} \left(\frac{T}{10^9\,\mathrm{K}}\right)^8.
\end{equation}
The fast processes scale as $T^6$, whereas the slow go as $T^8$.  To estimate the corresponding neutrino luminosity, we neglect the variation in neutrino emissivity with density and gravitational redshift and write $\Lnu\approx (4\pi R_c^3/3)\epsilon_\nu(\Tc)$, with core radius $R_c=11\ {\rm km}$, to obtain
\begin{eqnarray}
\label{e.LdUrca}
\LnudUrca &=& 6\times 10^{38}\,\Tc_8^6\,\ergpersec\\
\LnumUrca &=& 6\times 10^{30}\,\Tc_8^8\,\ergpersec .
\label{e.LmUrca}
\end{eqnarray}
The modified Urca cooling exceeds photon cooling in quiescence (Eq.~[\ref{e.Lph-heavy}]) for $\Tc_8 > 2.4$. 
Hence, modified Urca is not important during the outburst of \src.  If a slow cooling process (i.e., one $\propto \Tc^8$) were important for regulating the core temperature, it would need to be at least $10^3$ times stronger than modified Urca.

For a fast emission process, neutrino cooling exceeds radiative cooling at $\Tc_8 = 0.7$ for $\Lnu/\Tc_8^6 > 3.8\times 10^{33}\,\mathrm{erg\,s^{-1}}$, which is about $10^{-5}$ of the direct Urca luminosity.  This threshold is indicated by the vertical dotted line in Fig.~\ref{fig:phaseplot}.  We can rule out a neutrino emission as large as direct Urca, however, because the core neutrino luminosity cannot exceed the luminosity entering the core during outburst: $\Lnu < \Lin \approx 2\times 10^{35}\,\ergpersec$ (Fig.~\ref{fig:Lin}). As the core is heated and its temperature rises (assuming the heat capacity is low enough to give a large temperature change), the neutrino emissivity will eventually come into balance with the heating rate, and the core temperature will saturate. For direct Urca, the saturation temperature is much smaller than the inferred core temperature.  Setting $\Lnu < \Lin$ implies an upper limit to the emissivity of any fast neutrino process,
\begin{equation}
\frac{\epsilon_\nu^{\mathrm{fast}}}{(T/10^9\,\mathrm{K})^6} < 10^{23}\,\mathrm{erg\,cm^{-3}\,s^{-1}}
\left(\frac{\Tc_8}{0.7}\right)^{-6}\left(\frac{\Lin}{2\times 10^{35}\,\ergpersec}\right),
\end{equation}
which is about $10^{-3}$ of the direct Urca emissivity.
This limit is shown as the dark vertical line in Fig.~\ref{fig:phaseplot}.

Neutrino cooling during the outburst removes heat from the core; as a result, the core heat capacity can be below the limit in Eq.~(\ref{eq:Climit}). As $\Lnu \to \Lin$, the lower limit on heat capacity $C\to 0$. In this limit, very small values of $C$ are allowed because the core temperature reached during outburst is limited by the saturation value for which $L_\nu \approx \dot M \Qnuc$. This lower limit on $C$ as a function of $\Lnu$ is shown by the lower dark curve in Fig.~\ref{fig:phaseplot}. The lower limit on $C$ from Eq.~(\ref{eq:Climit}) remains valid until $\Lnu$ is within a factor of a few of the maximum allowed value. Such a large $\Lnu$ would, however, also produce a measurable decrease in the core temperature during quiescence.  We show next that current data already rule out such a large neutrino emissivity for \src.

\subsection{Cooling of the core during quiescence and future bounds on the core heat capacity}
\label{sec:timeevolution}


Cooling via neutrino losses during quiescence is potentially observable.  If we neglect $\Lph$ in Eq.~(\ref{eq:evolution-full}), then for a linear dependence on $\Tc$ for $C$ and $\Lnu\propto \Tc^\alpha$ the core temperature evolves over a time $\Delta t$ from $T_i$ to $T_f$ as
\begin{equation}\label{e.cooling-law}
\left(\frac{\Tc_i}{\Tc_f}\right)^{\alpha-2} - 1 = (\alpha-2)\frac{\Delta t}{\tau}
\end{equation}
where $\alpha = 6$ for fast neutrino cooling and
\begin{equation}\label{e.tau}
\tau = \frac{C\Tc}{\Lnu(\Tc)} \approx 3000\,\mathrm{yr}\;\frac{C_{38}\Tc_8}{L_{\nu,35}}
\end{equation}
is the cooling timescale.  For $\Delta t \ll \tau$, we may expand Eq.~(\ref{e.cooling-law}) and use Eq.~(\ref{e.tau}) to obtain 
\begin{equation}\label{e.C-Lnu}
\frac{C_{38}}{L_{\nu,35}} = \left(\frac{\Delta\Tc/\Tc}{0.3\%}\right)^{-1} \left(\frac{t_q}{10\,{\rm yr}}\right) \Tc_8^{-1}.
\end{equation}
Further monitoring of the quiescent temperature of \src\ can therefore either measure or limit $C/\Lnu$. If the core temperature remains constant, we will obtain a lower limit on $C$ as a function of $\Lnu$. 

Eq.~(\ref{e.C-Lnu}) shows that if neutrino losses during outburst are significant, then the temperature in quiescence should show a rapid decline. For example, setting $C=10^{36}\ {\rm erg\ K^{-1}}$ and $\Lnu=10^{35}\,\ergpersec$ gives $\Delta \Tc/\Tc \approx 30$\% over 10 years. A temperature change this large should be straightforward to observe. Indeed, the most recent temperature measurement for \src\ rules out temperature changes this large. Taking the two measurements of $\Teffinf=64.5\pm 1.8\ {\rm eV}$ and $64.4\pm 1.2\ {\rm eV}$ separated by 6 years \cite{Merritt2016}, we find the $1\sigma$ error in the slope is $0.4\ {\rm eV\ yr^{-1}}$. Assuming $\Tc\propto \Teff^{1.8}$ (appropriate for a heavy element envelope) gives a $2\sigma$ upper bound on the change in core temperature of $<13$\% over 6 years for \src\ (equivalent to $<20$\% over 10 years).  This lack of cooling excludes the dark shaded region in Fig.~\ref{fig:phaseplot}.  As a result, the lower limit on $C$ from  Eq.~(\ref{eq:Climit}) holds; indeed, for $\Lnu \lesssim \Lin \approx 2\times 10^{35}\,\ergpersec$, the lower limit on $C$ exceeds that of Eq.~(\ref{eq:Climit}).

If a decrease in core temperature is measured during quiescence, we can then determine $C/\Lnu$. Since we already have an upper limit on $\Lnu$, we would then have a corresponding upper limit on $C$. Setting $\Lnu = \Lin$ in Eq.~(\ref{e.C-Lnu}) gives
\begin{eqnarray}\label{eq:Cupper}
	C&<&10^{38}\ {\rm erg\ K^{-1}}\ \left({\Delta \Tc/\Tc\over 0.01}\right)^{-1} \left({t_q\over 10\ {\rm yr}}\right)\left({\Tc_8\over 0.7}\right)^{-1}\nonumber\\&&
    \left(\frac{\Lin}{2\times 10^{35}\,\ergpersec}\right).
\end{eqnarray}
The upper limit on core heat capacity we would obtain for a measured 5\% change in temperature over 10 years is indicated by the upper boundary of the light shaded region in Fig.~\ref{fig:phaseplot}. 
When combined with the lower limit from heating, we would then confine both the neutrino luminosity and heat capacity to a narrow range of possible values within the light shaded region.  Alternatively, tighter constraints on $\Delta\Tc/\Tc$ would increase the excluded area at the lower right in Fig.~\ref{fig:phaseplot}.

\subsection{The recurrence time and the core temperature}\label{s.repeated-outbutsts}

The duration of the quiescent period in \src\ is unknown. If the neutron star is quiescent for too brief a time for its core to completely cool, then it starts the subsequent outburst slightly hotter. This continues over repeated outbursts until the heat injected during outburst is radiated away by photons or neutrinos during quiescence. As a result, the core temperature at the end of the outburst is hotter than if it were heated only by that one outburst.

To illustrate this, we integrate Eq.~(\ref{eq:evolution-full}) in outburst neglecting $\Lph$ and $\Lnu$, so that the energy deposited into the core is $E = \Lin\tout$.  We then integrate over quiescence for a time $\trec-\tout$ with $\Lin\to0$, and apply the constraint that the temperature at the end of quiescence equal that at the start of the outburst.  We are assuming that $\Lnu \ll \Lin$; under these conditions, the core temperature at the end of outburst has a simple analytical form:
\begin{equation}\label{e.core-temperature-analytical}
\Tc = \left(\frac{2E}{C/\Tc_{8}}\right)^{1/2}
\left[1-\left(1+(\alpha-2)\frac{\trec}{\tau}\right)^{-2/(\alpha-2)}\right]^{-1/2}.
\end{equation}
Here $\trec$ is the outburst recurrence time and $\alpha$ is the temperature exponent for the cooling mechanism: if neutrino cooling dominates $\alpha = 6$ (8) for fast (slow) processes; if neutrino cooling is negligible then $\alpha = 2.2$ for cooling from radiative emission from the surface with a heavy element envelope (Eq.~[\ref{eq:Tc_heavy}]). The cooling timescale for fast neutrino losses is defined in Eq.~(\ref{e.tau}); for radiative cooling it is
\begin{equation}\label{eq:cooling-timescale}
\tau_\gamma = \frac{C\Tc}{\Lph(\Tc)} \approx 3\times 10^5\, C_{38} \Tc_8^{-1.2}\;\mathrm{yr}.
\end{equation}
In the limit $\trec\ll \tau$, Eq.~(\ref{e.core-temperature-analytical}) reduces to Eq.~(\ref{eq:Lbalance}).  For $t_\mathrm{r}\gg\tau$, the core temperature $\Tc$ becomes solely a function of the heat deposited during outburst and the neutron star core acts as a calorimeter with the temperature related to specific heat according to Eq.~(\ref{eq:Climit}).

\begin{figure}[htbp]
\includegraphics[width=\columnwidth]{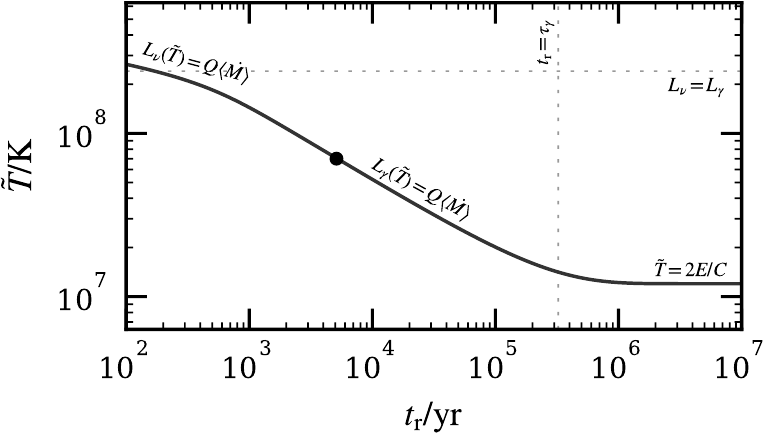} 
\caption{Core temperature $\Tc$ at the end of the outburst as a function of outburst recurrence time $\trec$.  The energy deposited into the core is $7.2\times10^{43}\,\mathrm{erg}$ and the specific heat is $C/\Tc_{8}=10^{38}\,\mathrm{erg\,K^{-1}}$. The light dotted vertical line indicates the radiative cooling timescale (Eq.~[\protect\ref{eq:cooling-timescale}]). For recurrence times longer than this, the core cools completely and the temperature at the end of the outburst obeys the simple relation, Eq.~(\ref{eq:Clim_intro}); for shorter recurrence times, the core temperature is set by balancing $E = \Lph \trec$ (cf.\ Eq.~[\protect\ref{eq:Lbalance}]) or by balancing $E = \Lnu \trec$ for sufficiently high $\Tc$, as indicated by
the horizontal dotted line marking where neutrino cooling (modified Urca; Eq.~[\protect\ref{e.LmUrca}]) equals radiative cooling.  At the inferred core temperature $\Tc = 7\times 10^7\,\mathrm{K}$, the recurrence time in this scenario is thus $5\,000\,\mathrm{yr}$, as indicated by the dark circle.
\label{fig:simple-model}
}
\end{figure}

The behavior of the core temperature $\Tc$ with recurrence time $\trec$ is illustrated in Figure~\ref{fig:simple-model}.  
For this plot, the heat deposited in an outburst is $7.2\times10^{43}\,\mathrm{erg}$, cf.\ eq.~(\ref{eq:E}) with an outburst time of $12\,\mathrm{yr}$; the specific heat has a linear temperature dependence with $C/\Tc_8 = 10^{38}\,\mathrm{erg\,K^{-1}}$; and the neutrino cooling is solely from modified Urca, Eq.~(\ref{e.LmUrca}).
The timescale for the core to cool is $\tau_\gamma \approx 3\times 10^5\,\mathrm{yr}$ (cf.\ Eq.~[\ref{eq:cooling-timescale}]).  For $\trec > \tau_\gamma$, the core temperature at the end of the outburst is set by the Eq.~(\ref{eq:Clim_intro}).  For shorter recurrence times, the core temperature is set by the need to radiate away the deposited energy in quiescence, Eq.~(\ref{eq:Lbalance}).  This is done by radiative emission from the surface. in the region marked $L_\gamma(\Tc) = Q\langle\dot{M}\rangle$.  At still shorter recurrence times, the core temperature is large enough that neutrino cooling becomes important (indicated by horizontal dotted line), and the temperature is set by $L_\nu(\Tc) = Q\langle\dot{M}\rangle$.

Given the inferred core temperature $\Tc = 7\times 10^7\,\mathrm{K}$, for this combination of specific heat and neutrino emission the recurrence time would be $\approx 5\,000\,\mathrm{yr}$, as indicated by the dark circle on the plot.  Different scenarios for $C$ and $\Lnu$ generate a family of curves $\Tc(\trec)$, and by setting $\Tc(\trec) = 7\times 10^7\,\mathrm{K}$ and solving for $\trec$, we can map out contours of constant $\trec$, as shown in Fig.~\ref{fig:phaseplot} (light grey curves).  Starting at the lower left of the plot, where both $C$ and $\Lnu$ are small, we are in the calorimeter regime (lower dark curve), and $\trec > \tau_\gamma$. Moving upwards on the plot to the larger $C$ while keeping $\Lnu$ small, we are in the regime $\Lph = Q\langle\dot{M}\rangle$, so that $\trec\approx 5\,000\,\mathrm{yr}$.  The moving from the upper left toward the upper right by increasing $\Lnu$, the recurrence time becomes progressively shorter so that $\Lnu = Q\langle\dot{M}\rangle$; as $\trec\to0$ we approach the limiting neutrino luminosity $\Lnu = \Lin$ (dark vertical line).

Another way to improve the lower limit on the heat capacity would be to use multiple outbursts from the same source. For example, a short recurrence time for \src\ would tightly constrain $\Lnu$ to lie close to its upper limit (vertical solid curve, Fig.~\ref{fig:phaseplot}), while limits on the variability would bound $C$ (\S~\ref{sec:timeevolution}).  If the core temperature were measured before and after an outburst, the change in temperature of the core due to the energy deposited during the outburst could be directly measured. The resulting constraint on the heat capacity would likely be much more constraining than our lower limit, which assumes that the core is very cold at the start of the outburst. A complication in doing this is that the envelope composition will most certainly be different in the two quiescent periods \cite{Brown2002}. This could perhaps be resolved by modeling the shape of the quiescent cooling curve, as we have done here for \src\ (Sec.~\ref{s.limit-core-C}). One source for which this could be attempted in the near future is \mxb, which recently went into outburst again after more than 14 years in quiescence \cite{SanchezFernandez2015}.

\section{Conclusions}

We have shown that observations of the temperature of accreting neutron stars in quiescence provide a lower limit to the heat capacity of the neutron star core. This limit is derived by assuming that the neutron star core cools completely between outbursts, opposite to the usual assumption that the core is in long term equilibrium, used to constrain the core neutrino emissivity. The core is then a calorimeter that can be used to determine the heat capacity given the energy deposited and final core temperature. The main uncertainty in deriving the lower limit is the envelope composition, which can change the inferred core temperature by a factor of $2$--$3$. However, the envelope composition is constrained by the shape of the cooling curve; in particular we show that \src\ is best fit with a heavy element envelope (as also recently pointed out by \cite{Ootes2016}). The lower limits to the core heat capacity for the sources \src, \mxb, and \xtej\ are in the range $C\gtrsim 2$--$5\times 10^{36}\,\Tc_8\ {\rm erg\ K^{-1}}$, where $\Tc_8$ is the core temperature in units of $10^8\ {\rm K}$. This is a factor a $2$--$3$ below the heat capacity expected from electrons, which set the heat capacity when the nucleons are superfluid in the core. This limit rules out a large fraction of the core being made up of a CFL phase. We have also shown that continued observations in quiescence can strengthen the lower limit and, if cooling in quiescence is detected, provide a complimentary upper limit on the core heat capacity. Long timescale observations of quiescent neutron stars provide a new way to constrain the unknown composition of dense matter in neutron star cores.
\vspace{1\baselineskip}

{\itshape Acknowledgments}--- We thank A.~Deibel for comments on the manuscript. We thank the International Space Science Institute (ISSI) in Bern for hospitality and support of an International Team on neutron star crusts, through which this work began. AC is supported by an NSERC Discovery Grant.  EFB is supported by the US National Science Foundation under grant AST-1516969. DP is partially supported by the Consejo Nacional de Ciencia y Tecnolog{\'\i}a with a CB-2014-1 grant $\#$240512. CJH and FJF are supported by DOE grants DE-FG02-87ER40365 (Indiana University) and DE-SC0008808 (NUCLEI SciDAC Collaboration). SR was supported by the DOE Grant No.\ DE-FG02-00ER41132 and by the National Science Foundation under Grant No.\ PHY-1430152 (JINA Center for the Evolution of the Elements).  

\bibliography{bibtex}

\end{document}